\pdfoutput=1
\documentclass[twocolumn]{aastex62}
\usepackage{natbib}
\usepackage{graphicx}
\usepackage {amsmath,amssymb,eqnarray}
\usepackage{enumerate}
\usepackage{enumitem}
\usepackage{mathtools}

\newcommand {\hi} {{\rm H}{\textsc i}}

\newcommand {\htwo} {{\rm H}_2}
\newcommand{\dHI}{D_{{\rm HI}}}
\newcommand{\rHI}{R_{{\rm HI}}}
\newcommand{\mHI}{M_{{\rm HI,tot}}}
\newcommand{\fHI}{f_{{\rm HI,tot}}}
\newcommand{\fHtwo}{f_{{\rm H_2}}}
\newcommand{\mHtwo}{M_{{\rm H_2}}}
\newcommand{\mHIin}{M_{{\rm HI,in}}}
\newcommand{\fHIin}{f_{{\rm HI,in}}}
\newcommand{\mHIinpred}{M_{{\rm HI,in,pred}}}
\newcommand{\mHIoutpred}{M_{{\rm HI,out,pred}}}

\newcommand{\Sigmastd}{\tilde{\Sigma}_{\rm H}}
\newcommand{\SHI}{\Sigma_{\rm HI}}
\newcommand{\SHIin}{\Sigma_{\rm HI,in}}

\newcommand{\rd}{r_{\rm disc}}
\newcommand{\fc}{R_{\rm mol}^c}

\begin{document}
\title{ \lowercase{x}GASS: HI fueling of star formation in disk-dominated galaxies }

\author{Jing Wang}
\affiliation{ Kavli Institute for Astronomy and Astrophysics, Peking University, Beijing 100871, China}

\author{Barbara Catinella}
\affiliation{International Centre for Radio Astronomy Research (ICRAR), University of Western Australia, Crawley, WA 6009, Australia}
\affiliation{Australian Research Council, Centre of Excellence for All Sky Astrophysics in 3 Dimensions (ASTRO 3D), Australia}

\author{Am{\' e}lie Saintonge}
\affiliation{Department of Physics \& Astronomy, University College London, Gower Place, London WC1E 6BT, UK}

\author{Zhizheng Pan}
\affiliation{Purple Mountain Observatory, Chinese Academy of Sciences, 2 West-Beijing Road, Nanjing 210008, People's Republic of China }

\author{Paolo Serra}
\affiliation{INAF, Osservatorio Astronomico di Cagliari, Via della Scienza 5, 09047 Selargius, CA, Italy }

\author{Li Shao}
\affiliation{National Astronomical Observatories, Chinese Academy of Sciences, 20A Datun Road, Chaoyang District, Beijing, China}

\begin{abstract}
We introduce a method to estimate the $\hi$ mass within the optical radius of disk galaxies from integrated HI spectra, with an uncertainty of 0.09 dex. We use these estimates to study how inner $\hi$ fuels star formation in late-type disk galaxies.
We find that star formation rate (SFR) at a given stellar mass ($M_*$) is well correlated with the inner $\hi$ surface density ($\SHIin$) and inner $\hi$ mass-to-stellar mass ratio. For the massive ($M_*>10^{10} M_{\odot}$) disk galaxies, higher SFR at a given stellar mass is also related with higher efficiency of converting inner $\hi$ to molecular gas, but no such correlation is found for the total $\hi$ mass. The highest $\SHIin$ and the fastest depletion of the total neutral gas within the optical disks are found in the most compact and star-forming disk galaxies at a given stellar mass. These results highlight the important role of inner $\hi$ as an intermediate step of fueling star formation in disk galaxies. 
\end{abstract}

\keywords{spiral galaxies, neutral atomic gas}

\section{Introduction}
\label{sec:introduction}
Galaxies in the low and high redshift Universe are distributed in remarkably regular patterns in the parameter space of star formation rate ($SFR$) versus stellar mass ($M_*$) \citep{Noeske07, Schiminovich07, Elbaz11, Whitaker12, Speagle14, Whitaker14, Tacconi18}. Massive ($M_*>10^9~M_{\odot}$) star-forming galaxies are concentrated onto a tight relation between $\log SFR$ and $\log M_*$, with a slope less than unity, and a typical scatter of 0.3-0.4 dex in SFR throughout the relation \citep{Noeske07, Speagle14, Whitaker14}. This tight relation is referred to as the star forming main sequence (SFMS). The less star-forming galaxies show a broad range of SFR at a fixed $M_*$ below the SFMS. 

Theories predict that the scatter of the SFMS is caused by the oscillation of galaxies around the SFMS with time, which is driven by a balance between cold gas replenishment and depletion in their inner disk \citep{Dekel14, Zolotov15, Tacchella15}. If replenishment is quicker than depletion, the central gas density increases, the SFR rises above the SFMS and the central stellar mass concentration goes up. This process continues until the very high central gas surface density triggers vigorous star formation followed by strong feedback that quickly depletes the gas. This process of central gas, SFR and stellar density building up until the onset of quick depletion is called compaction \citep{Dekel14, Zolotov15, Tacchella15}. At the end of compaction, SFR drops below the SFMS, until replenishment of gas brings the galaxy back to a new circle of compaction. Supporting observational evidence includes that passive galaxies start to be abundant when the central stellar density reaches a threshold value \citep{Fang13, Woo15, Tacchella16b, Mosleh17, Whitaker17}, compact star-forming and compact passive galaxies have similar mass, kinematics and morphology \citep{Barro13, Barro14, Bruce14, Nelson14, Williams14, Barro17}, and galaxies quench inside-out \citep{Tacchella15, GonzalezDelgado16, Barro17, Brennan17, Belfiore17, Ellison18}.

For galaxies to stay on the SFMS, their star formation needs to be sustained by gas accretion \citep{Kennicutt98, Putman17}. The accreted gas need go through a whole process of cooling out of the circum-galactic medium \citep[CGM][]{White91}, in-falling and settling into a dynamically and thermally cool atomic hydrogen ($\hi$) disk, and further cooling and condensing into molecular ($\htwo$) clouds \citep{Krumholz12} in the inner disks, before finally fueling the star formation \citep{Bigiel08, Schruba11}. Thus, $\hi$ is a necessary intermediate phase for the fueling (sustaining) of star formation. In the more distant Universe, the phase of $\hi$ might be less important, for the neutral gas seems to be dominated by the molecular phase (\citealp{Tacconi13, Tacconi18}; but see also e.g. \citealp{Cortese17, Decarli19}). But in the local Universe, the conversion of $\hi$ to $\htwo$ is much less efficient and $\hi$ is a significant component of the neutral gas ($\sim$80\% globally and 50\% in the stellar disks for star-forming galaxies, \citealt{Catinella18} and this paper). Hence the $\hi$ abundance and its connection to the $\htwo$ phase are important details to quantify observationally in order to constrain models of galaxy evolution. 

At the present, a most efficient way to directly observe neutral hydrogen gas for a statistically, significantly large sample of galaxies is through single-dish radio telescopes \citep{Saintonge17, Catinella18, Haynes18}. The integrated mass and velocity widths of the neutral gas obtained from single-dish surveys, has greatly advanced our understanding of galaxy population and evolution. The $\hi$ and $\htwo$ mass fractions (gas mass over the stellar mass) are found to be correlated with the colour, specific SFR and effective stellar mass surface density of galaxies \citep{Catinella10, Saintonge11, Tacconi13, Saintonge17, Catinella18}. A high  $\hi$-richness is also related with newly formed outer disks  with low gas-phase metallicities \citep{Moran10, Moran12, Carton15}, and an excess of young stars \citep{Wang11}, a significant fraction of which are formed in bursts \citep{Huang13}.  $\hi$-rich galaxies further tend to be in environments of low local densities and low dark matter halo masses \citep{Fabello12, Catinella13, Hess13, Brown17}. The depletion time of $\hi$ ($\mHI/SFR$) is typically 3 $Gyr$ for different types of massive galaxies \citep{Schiminovich10}, but can be close to the Hubble time for low-mass and low-SFR galaxies \citep{Saintonge17}. The star-forming efficiency of $\htwo$ (characterized as $SFR/\mHtwo$) is strongly correlated with the specific SFR (sSFR) and stellar surface density of galaxies \citep{Saintonge11, Saintonge12, Tacconi13, Huang14, Genzel15, Huang15}. Studies of the distribution of galaxies in the $SFR$-$M_*$ space found that the $\hi$ and $\htwo$ mass fractions are not only correlated with the extent of SFR deviating from the SFMS, but also anti-correlated with the stellar mass along the SFMS  \citep{Saintonge16}. The former correlation is strengthened when the galaxies have high central stellar compactness \citep{Wang18}.

However, the lack of spatial information has become a major limitation to linking the observed neutral gas properties to physical processes. Especially, $\hi$ is typically more radially extended than the stellar disk in star-forming galaxies \citep{Swaters02, Wang13}, and the high angular momentum $\hi$ in the outer region cannot effectively feed the star formation which is largely concentrated in the stellar disks \citep{Bigiel10, Yim16, Wang17, Yildiz17}. In the future, the Square Kilometer Array (SKA) and its pathfinders will provide high-resolution $\hi$ images for a large area of the sky \citep{deBlok15, Staveley-Smith15}, and will finally solve this problem. But before that, we already have the knowledge to roughly estimate the spatial distribution of $\hi$ gas from the integrated $\hi$ data, at least for late-type disk galaxies. The radial distributions of $\hi$ are self-similar in the outer regions of late-type galaxies when normalized to a characteristic radius $R_{\rm HI}$, defined as the semi-major axis of the 1 $M_{\odot}~pc^{-2}$ isophote \citep[][W16 hereafter]{Swaters02, Wang14, Wang16}. Partly due to this similarity, there is a tight linear relation between the $\hi$ mass $\mHI$ and $R_{\rm HI}$, with a scatter of only 0.06 dex for estimates of $R_{\rm HI}$ based on the relation \citep{Broeils97, Verheijen01, Swaters02, Noordermeer05, Wang14, Martinsson16, Wang16}. Combining these two facts, it is possible to guess$\slash$predict the outer part of the $\hi$ surface density radial profile of late-type galaxies given the $\hi$ mass, and hence divide the integrated $\hi$ mass into inner and outer parts. In this paper, we explore the feasibility and application of such a method. In particular, we study how the predicted $\hi$ located within the inner stellar disks will alter or reinforce our established view of $\hi$ fueling the star formation of star-forming galaxies. 

The paper is organized as follows. We introduce the main sample selected from xGASS and xCOLD GASS, and a validation sample selected from $\hi$ interferometric surveys of nearby galaxies in Sec. 2. In Sec. 3, we describe the method of predicting the $\hi$ mass within optical radius from the total $\hi$ mass, and use the validation sample to justify and calibrate the method. In Sec. 4, we apply the method to the main sample and analyze how $\hi$ masses and densities within the optical radius vary in the space of $SFR$ versus $M_*$ of galaxies. We discuss the results in Sec. 5. We adopt a $\Lambda$CDM cosmology with $\Omega_{m}=0.3$, $\Omega_{\lambda}=0.7$ and $h=0.7$. The initial mass function of \citet{Chabrier03} has been assumed for stellar mass and SFR estimates. 

\section{Data}
\label{data}

\subsection{The main sample}
\label{sec:main_sample}
This study is based on the xGASS (extended GALEX Arecibo SDSS Survey) representative sample \citep{Catinella18} of 1179 galaxies selected by the stellar mass ($M_*>10^9~M_{\odot}$) and redshift ($0.01<z<0.02$ for $M_*<10^{10}~M_{\odot}$ and $0.025<z<0.05$ for $M_*>10^{10}~M_{\odot}$). The single-dish $\hi$ data were obtained with the Arecibo telescope, mostly in the GASS \citep{Catinella10, Catinella13} and GASS-low \citep{Catinella18} surveys, and complemented by data from ALFALFA \citep{Giovanelli05, Haynes11} and Cornell $\hi$ digital archive \citep{Springob05}. Flags were provided for each $\hi$ spectrum to indicate the detection quality and possible confusion. Unless specifically noted, we do not account for the helium in the $\hi$ and $\htwo$ data used in this paper.

Additional multi-wavelength information was collected from the public databases. Spectroscopic and photometric measurements are taken from SDSS-DR7 \citep{Abazajian09}, including the redshift, the radius $r_{50}$ and $r_{90}$ (radius that enclose 50\% and 90\% of the total flux, respectively), and radial distribution of surface brightness in the optical bands $u$, $g$, $r$, $i$ and $z$. We derive the average stellar surface densities within the central 1 kpc through interpolating the radial profiles of surface brightness in the $g$ and $r$ bands, and converting the $r$-band surface brightness to stellar mass surface densities with $g-r$ dependent $M_*$-to-light ratio from \citet{Zibetti09}. Additional estimates of galactic properties are taken from the MPA$\slash$JHU catalogue \citep{Kauffmann03}, including stellar mass $M_*$ for all the galaxies and gas-phase metallicity $O/H$ for the galaxies with strong emission lines. 

Star formation rates (SFR) were estimated based on the combination of NUV \citep[from GALEX,][]{Morrissey07} and MIR \citep[from WISE,][]{Wright10} luminosities \citep{Janowiecki17}. There are 532 galaxies from xGASS followed up by the IRAM 30 m telescope to obtain the CO(1-0) emission line fluxes \citep[xCOLD GASS,][]{Saintonge11, Saintonge17}. The conversion factor $\alpha_{CO}$ was derived based on the metallicity and offset from the SFMS \citep{Accurso17}, to convert the CO(1-0) fluxes to the $\htwo$ masses.

We select the disk-like, $\hi$ detected galaxies from the xGASS representative sample by requiring $r$-band light concentration $r_{90}/r_{50}<2.7$, reasonable $\hi$ detection quality ($HI\_FLAG=$1 or 2 \footnote{If we remove the $HI\_FLAG=$2, marginal detections, the trends presented in this paper do not change}), and no significant $\hi$ confusion ($HIconf\_flag=$0). The selection on $r_{90}/r_{50}$ is to select (late-type) disk-like galaxies which have self-similar radial distributions of $\hi$ in the outer regions, and no significant stellar bulges which may affect the partition of neutral hydrogen into atomic and molecular phases in the inner regions. These properties are the basis of the method which we are going to use to estimate the $\hi$ mass within the optical disks (Sec.~\ref{sec:method}).
The selection results in a {\it xGASS-disk sample} of 447 galaxies. Among it, 179 have CO detections from xCOLD GASS, which make the {\it main sample} of this paper.

\subsection{The validation sample}
\label{sec:test_sample}
We estimate the $\hi$ mass within the optical $r_{90}$ ($r$-band) of galaxies, based on the integrated $\hi$ mass ($\mHI$) and optical photometric measurements. We will calibrate and test the methods, against a validation sample of nearby galaxies which have well-resolved and sensitive $\hi$ images (naturally weighted). 

The validation sample (VS) includes galaxies from THINGS \citep{Walter08}, WHISP-Sc \citep{Swaters02} and LVHIS \citep{Koribalski18}. The stellar masses were calculated in \citet{Wang17}, based on Spitzer IRAC 3.6 and 4.5 $\mu$m luminosities for THINGS, the B \citep[from SIMBAD,][]{Wenger00} and R-band luminosities \citep[from][]{SwatersBalcells02} for WHISP-Sc, and the WISE 3.4 and 4.6 $\mu$m luminosities for LVHIS. $r_{90}$ were measured from 3.6 $\mu$m images for the THINGS galaxies \citep{Leroy08}, and 3.4 $\mu$m images for the LVHIS galaxies \citep{Wang17}. Only $r_{80}$ is available for WHISP-Sc  \citep{SwatersBalcells02}, so we approximate $r_{90}=1.1r_{80}$ for these galaxies. SFRs of THINGS galaxies were derived by combining
24 $\mu$m luminosities from SINGS \citep{Kennicutt03} and FUV luminosities from NGS \citep{GildePaz07}.  SFRs of LVHIS galaxies were estimated by combining the GALEX FUV and WISE 22 $\mu$m luminosities in \citet{Wang17}. We use the same pipeline as \citet{Wang17} to estimate SFR for the WHISP-Sc galaxies. 

We select the galaxies with $M_*>10^9~M_{\odot}$ and $r_{90}>2.5B_{maj}$. The selection on $M_*$ is to ensure a similar $M_*$ range as the main sample. The selection on size is to ensure reliable measurements of the $\hi$ mass within $r_{90}$.
 We further select the galaxies with no significant missing-flux problems, by requiring $\dHI<7\arcmin$ for THINGS galaxies (so that the interferometric $\hi$ fluxes are consistent with the single-dish ones within a scatter of $\sim$10\% after applying the selection criteria \citealt{Walter08}), and $\dHI< 6.7\arcmin$ for WHISP-Sa galaxies (suggested by \citealt{Swaters02}). VS includes 11, 29 and 10 galaxies from THINGS, WHISP-Sc, and LVHIS respectively, in total 50 galaxies.

\section{Estimating HI mass within the optical $r_{90}$ of disk-like galaxies}
\label{sec:method}
We present a method to estimate the $\hi$ mass within the optical $r_{90}$, $\mHIinpred$, based on $\mHI$ and other optical properties. 
We compare $\mHIinpred$ with the real measurements $\mHIin$ to assess the method. 
The method discussed below is independent of the selected radius $r_{90}$, which can be replaced by other types of radius (e.g. $r_{25}$ where the optical band isophotes reach a surface brightness of 25 mag arcsec$^{-2}$) in future applications. 

\subsection{Method: median $\SHI$ observed for galaxies}
\label{sec:method1}
\begin{figure}
\centering
\includegraphics[width=\linewidth, angle=0]{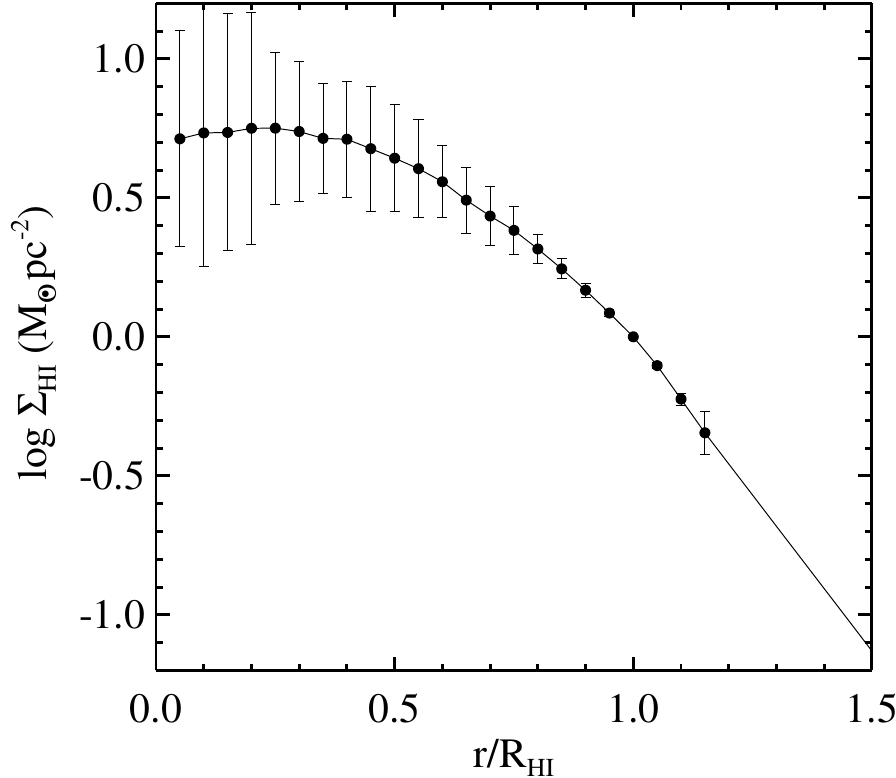}
\caption{The W16 $\SHI$ profile, obtained as the median of the $\SHI$ profiles of 168 spiral and dwarf galaxies derived in W16. The directly derived profile extends to 1.15$\rHI$, and has been extrapolated to 1.5$\rHI$. }
\label{fig:medianprof}
\end{figure}

This method makes use of two observational facts: 1) galaxies lie on a remarkably tight $\dHI-\mHI$ relation \citep[][W16]{Swaters02}; 2) the $\SHI$ profiles of late-type galaxies show homogeneous shapes in the outer regions, when the radius is normalized to $\rHI$ ($0.5\dHI$, \citealt{Wang14}, W16). 
We take the median $\hi$ surface density profile from W16, which has the radius normalized to $\rHI$, and was derived from a sample of 168 nearby spiral and dwarf galaxies which have $\rHI>3B_{maj}$, where $B_{maj}$ is the major axis of the synthesis beam. The original median profile extends to a maximum radius of 1.15$\rHI$, we extrapolate it out to 1.5$\rHI$ assuming an exponential outer disc with a scale-length of 0.2$\rHI$ \citep[][W16]{Wang14}.
 We refer to the extrapolated profile as the W16 $\SHI$ profile hereafter, and show it in Fig.~\ref{fig:medianprof}. 

\begin{figure*}
\centering
\includegraphics[width=\linewidth, angle=0]{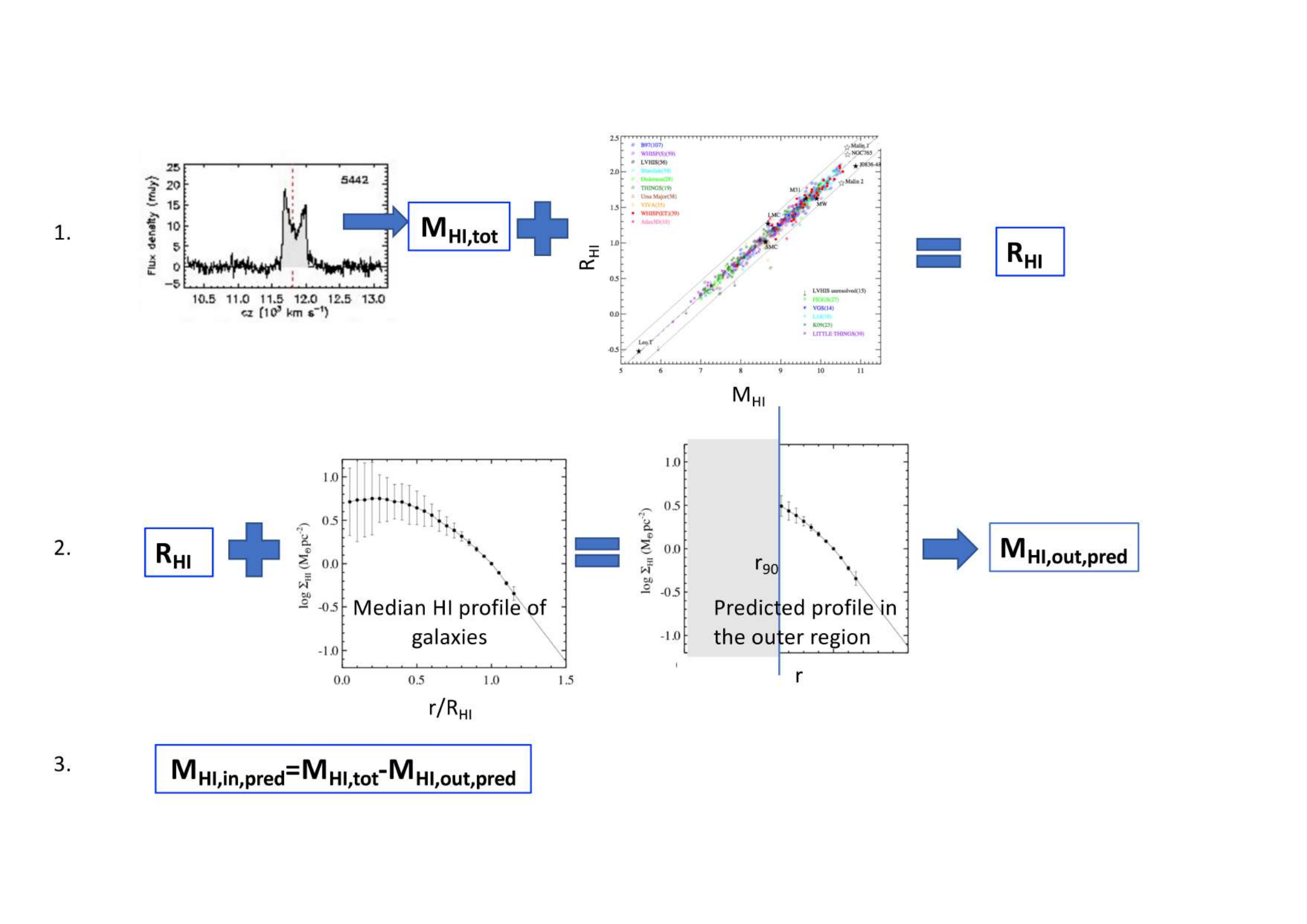}
\caption{Three steps that estimate $\mHIinpred$ based on $\mHI$ of a galaxy.}
\label{fig:procedure}
\end{figure*}

We describe the procedure of estimating $\mHIinpred$ below, and also demonstrate it in Figure~\ref{fig:procedure}.

For each given $\mHI$ of a galaxy, we estimate the radius $\rHI$ based on the $\dHI-\mHI$ relation. Then we estimate $\mHIoutpred$, the $\hi$ mass between $r_{90}$ and 1.5$\rHI$ with the following procedure: 

\begin{enumerate}
\item When $1.5\rHI>r_{90}$, we scale the radius of the W16 $\SHI$ profile by $\rHI$, and hence obtain a ``predicted $\SHI$ profile'' for the galaxy.
 Then we cumulate the predicted $\SHI$ profile between $r_{90}$ and 1.5$\rHI$ to estimate $\mHIoutpred$. 
 
 \item When $1.5\rHI<r_{90}$, $\mHIoutpred=0$.
\end{enumerate}

Finally, $\mHIinpred=\mHI-\mHIoutpred$.  
 
\subsection{Justification of the method}
\label{sec:justification}
We justify the performance of the method by quantifying the difference between the real and predicted amount of $\hi$ within $r_{90}$, $\log~M_{\rm HI,in,pred}/M_{\rm HI,in}$ and $\Sigma_{\rm HI,in,pred}-\Sigma_{\rm HI,in}$. We have subtracted 0.04 dex from the direct estimates of $M_{\rm HI,in,pred}$, to minimize the scatter and median offset from real measurements in VS. The -0.04 dex offset is likely due to wiggles in the $\SHI$ profile of individual galaxies, which are missed by the median $\SHI$ profile of W16 and cause $\mHIoutpred$ to under-estimate $M_{\rm HI,out}$. This correction will also be added when the method is applied to the main sample. In this sense, VS serves not only as an assessment sample, but also as a calibration sample for our method. 

\begin{figure*}
\includegraphics[width=9cm, angle=0]{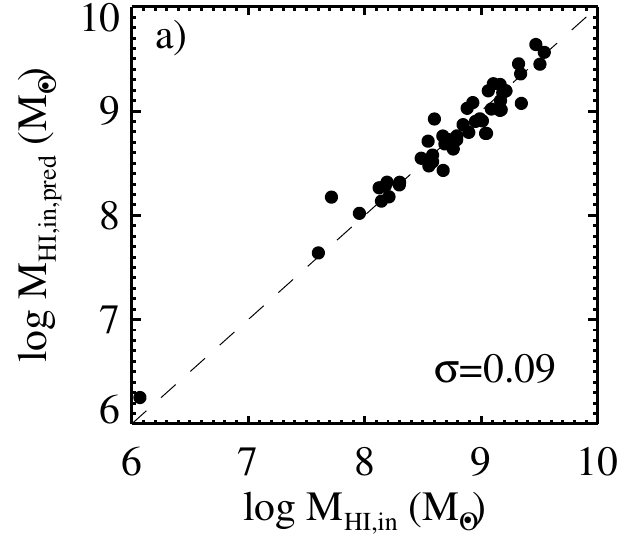}

\includegraphics[width=16cm,angle=0]{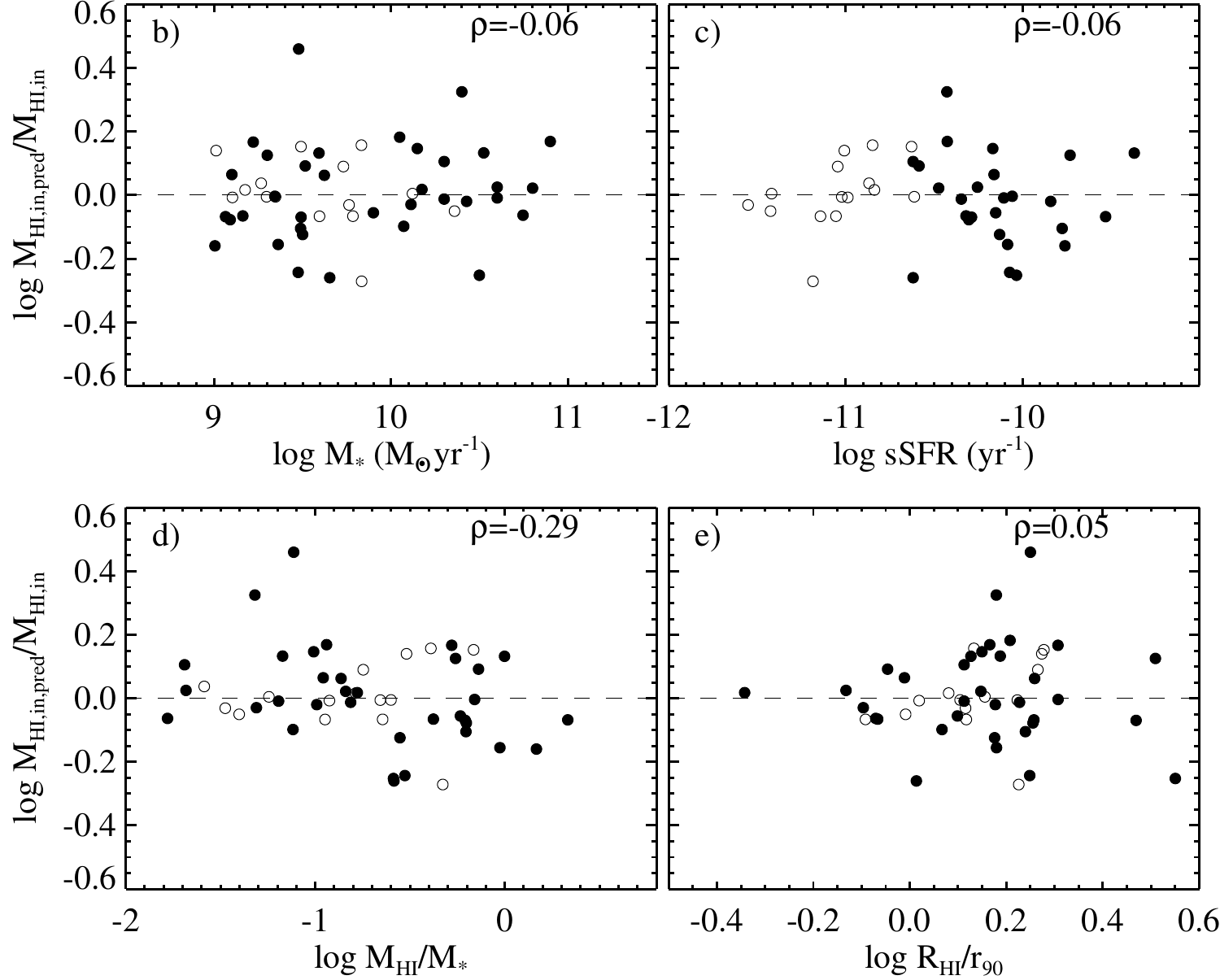}
\caption{Comparison between $M_{\rm HI,in,pred}$ and $\mHIin$ of the VS galaxies. In panel a, the dashed line is the $1:1$ line, and the scatter around the line is denoted in the corner. We also show the dependence of $\log~M_{\rm HI,in,pred}/M_{\rm HI,in}$ on $M_*$, $\log~sSFR$, $\mHI/M_*$ and $\rHI/r_{90}$ in panels b-e. In these panels, the dashed lines mark the position of $\log~M_{\rm HI,in,pred}/M_{\rm HI,in}=0$, and the Pearson correlation coefficients are denoted at the corners. The open circles mark the VS galaxies that do not overlap with the main sample (squares in Fig.~\ref{fig:dmhi_sfms}). }
\label{fig:dmhi_correlate}
\end{figure*}

Panel a of Fig.~\ref{fig:dmhi_correlate} shows the good correlation and small offset between $M_{\rm HI,in,pred}$ and $\mHIin$.
The scatter of $\log~M_{\rm HI,in,pred}/M_{\rm HI,in}$ and $\Sigma_{\rm HI,in,pred}-\Sigma_{\rm HI,in}$ are 0.09 dex and 0.6 $M_{\odot}~kpc^{-2}$ \footnote{We note that we have built a relatively small VS to match the $M_*$ range of the main sample. We also test the method with all the resolved late-type galaxies from W16, excluding the VIVA sample of galaxies in the Virgo cluster. We find $\log~M_{\rm HI,in,pred}/M_{\rm HI,in} \sim 0.02\pm$0.11 dex, and $\Sigma_{\rm HI,in,pred}-\Sigma_{\rm HI,in} \sim 0.15\pm$1.64 $M_{\odot}$. The relatively large uncertainty in estimating $\SHIin$ is mainly due to the fact that the optical radius of dwarf irregular galaxies tend to be small and reach the non-exponential part of the W16  $\SHI$ profile where the uncertainty is large. }. VS has an average $\log \mHIin/M_{\odot}$ of 6.59$\pm0.24$ dex, and average $\Sigma_{\rm HI,in}$ of 4.2$\pm$2.2 $M_{\odot}~kpc^{-2}$. Hence the uncertainty of both estimates are 2-3 times less than the scatter of the real values within the sample. 

Panels b-e of Fig.~\ref{fig:dmhi_correlate} show that the scatter of $\log~M_{\rm HI, in,pred}/M_{\rm HI, in}$ for the VS galaxies does not significantly depend on $M_*$, $sSFR$, $\mHI/M_*$, or $\rHI/r_{90}$. The Pearson correlation coefficients suggest a weak anti-correlation of $\log~M_{\rm HI, in,pred}/M_{\rm HI, in}$ with $\mHI/M_*$ ($\rho \sim0.29$). This weak anti-correlation does not significantly affect the interpretation of trends presented below in this paper, as long as the trends are $M_{\rm HI, in,pred}$ related properties being positively correlated with $\mHI/M_*$ (because if so the real trend should be even stronger than observed).

\begin{figure}
\centering
\includegraphics[width=8cm,angle=0]{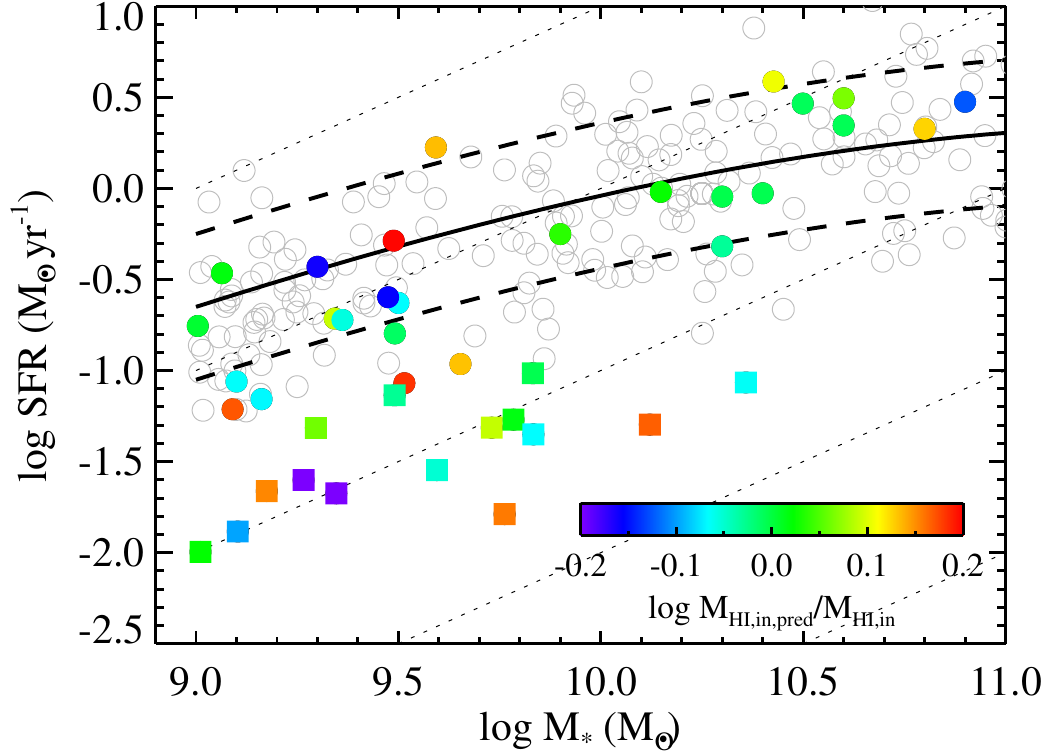}
\vspace{0.2cm}
\caption{The distribution of VS galaxies in the space of SFR versus $M_*$. The colors indicate $\log~M_{\rm HI, in,pred}/M_{\rm HI, in}$ of VS galaxies (colored symbols). Colored squares mark the VS galaxies that do not overlap with the main sample (open, grey circles). The solid and dashed curves mark the mean position of the SFMS \citep{Saintonge16} and the 0.4 dex deviations. The dotted lines mark positions of constant sSFR, with separations of 1 dex.  }
\label{fig:dmhi_sfms}
\end{figure}

Fig.~\ref{fig:dmhi_sfms} shows that $\log~M_{\rm HI, in, pred}/M_{\rm HI, in}$ do not show significant dependence on positions in the space of $SFR$ versus $M_*$. It also shows that VS galaxies do not exactly overlap with main sample galaxies in the space of $SFR$ versus $M_*$. But we can see from both Fig.~\ref{fig:dmhi_sfms} and the filled circles in Figure~\ref{fig:dmhi_correlate}, that the overlapping and non-overlapping VS galaxies do not have significantly different uncertainties in $M_{\rm HI, in, pred}$ with respect to $M_{\rm HI, in}$. 

We also find that $\mHIin$ estimated with this observation motivated method is as good as (or even better than) estimates produced by complex theoretical model (details in Appendix 1). But this method requires much fewer inputs (only $\mHI$ and $r_{90}$) and relies on much fewer assumptions than those models.

These validations lend us confidence that the method can be applied to the main sample to investigate the average $\mHIin$ and $\SHIin$ of galaxies along and around the SFMS. For simplicity, we refer to $\mHIinpred$ and $\Sigma_{\rm HI,in,pred}$ as $\mHIin$ and $\SHIin$ for the analysis of main sample galaxies hereafter. 

\subsection{Scaling relation of $\fHIin$ in the main sample}
\begin{figure*}
\centering
\includegraphics[width=18cm,angle=0]{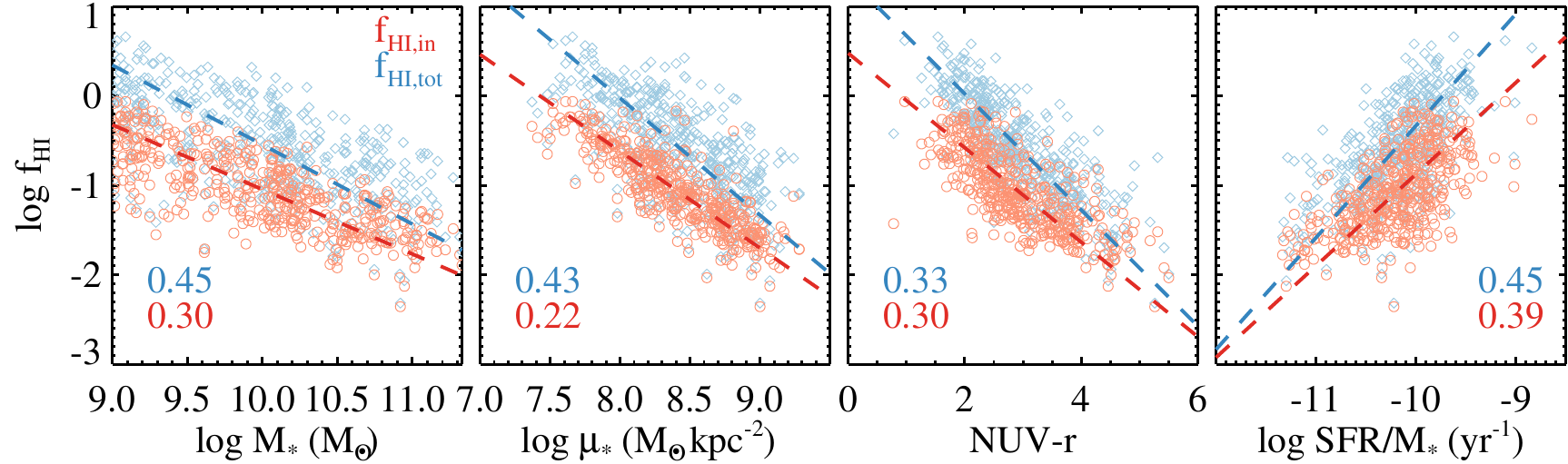}
\vspace{0.2cm}
\caption{Scaling relations of $\fHI$ and $\fHIin$. The dashed lines show the bisector linear fits. The scatters of $\fHI$ and $\fHIin$ relations are shown in each panel. Galaxies from the xGASS-disk sample are plotted. }
\label{fig:scaling}
\end{figure*}

We apply our method of deriving $\mHIin$ to the main sample of disk galaxies from xGASS. 
$\log \mHIin/\mHI$ ranges from -0.83 to -0.12 dex (15-76\% in percentage, the 5 and 95 percentiles), with a width roughly 41\% of the distribution width of $\fHI$ ($=\mHI/M_*$), and a median value of -0.53 dex (29\%). The relatively wide range and low median value of $\mHIin/\mHI$ lend support to the necessity of considering the inner $\hi$ when studying the star forming status of galaxies. 
We present in Figure~\ref{fig:scaling}, the scaling relations between $\fHIin=\mHIin/M_*$ and $M_*$, $\mu_*$ (the average stellar mass surface density with the $z$-band $r_{50}$), $NUV-r$ and the specific $SFR$ ($sSFR=SFR/M_*$). The trends are similar to the scaling relations of $\fHI$ \citep{Catinella18}: galaxies tend to have higher $\fHIin$ at lower $M_*$, lower $\mu_*$, bluer $NUV-r$ and higher $sSFR$. The slopes of $\fHI$ and $\fHIin$ relations differ most when the x-axis is $NUV-r$ or $sSFR$, because $\fHI$ and $\fHIin$ are close to each other when the $\hi$ disk shrinks into the stellar disks, i.e., when $NUV-r$ is red and $sSFR$ is low. The scatter of the $\fHIin$ relations is always smaller than that of the corresponding $\fHI$ relations, implying a closer link of $\mHIin$ with the stellar disks than $\mHI$.

\section{$\mHIin$ and $\SHIin$ of galaxies in the space of $SFR$ versus $M_*$}
\label{sec:SFMS}
This paper investigates the relation between the neutral gas reservoir of late-type
galaxies and the bulk of the star formation which takes place in the stellar disk.
We only consider the processes after the $\hi$ is accreted onto the disk, from either the circum-galactic medium or satellite galaxies. The $\hi$ in an  $\hi$-rich galaxy is radially more extended than the optical disk, and needs to flow into the stellar disk where star formation can happen efficiently. 
The efficiency for the whole $\hi$ reservoir to become available for star formation within the stellar disk can be quantified as $\mHIin/\mHI$. The mass of this inner $\hi$ reservoir can be quantified as $\mHIin$ or $\fHIin=\mHIin/M_*$. In addition, we use $\SHIin$, the average surface density of $\hi$ within $r_{90}$, for densities are physically more meaningful parameters to describe star forming activity than masses. The $\hi$ in the stellar disk will cool to form the molecular hydrogen and then stars. The efficiency of the former process can be quantified as $\mHIin/\mHtwo$. 

Finally, we use $\mHtwo+\mHIin$ to indicate the total reservoir of material directly available for forming stars on the stellar disks. We define the depletion time for this immediate gas reservoir $t_{\rm dep,in}=(\mHtwo+\mHIin)/SFR$.

We investigate these neutral gas related parameters in the space of SFR versus $M_*$.

\subsection{Trends as a function of $M_*$ along the SFMS}
\begin{figure*}
\centering
\includegraphics[width=5.8cm,angle=0]{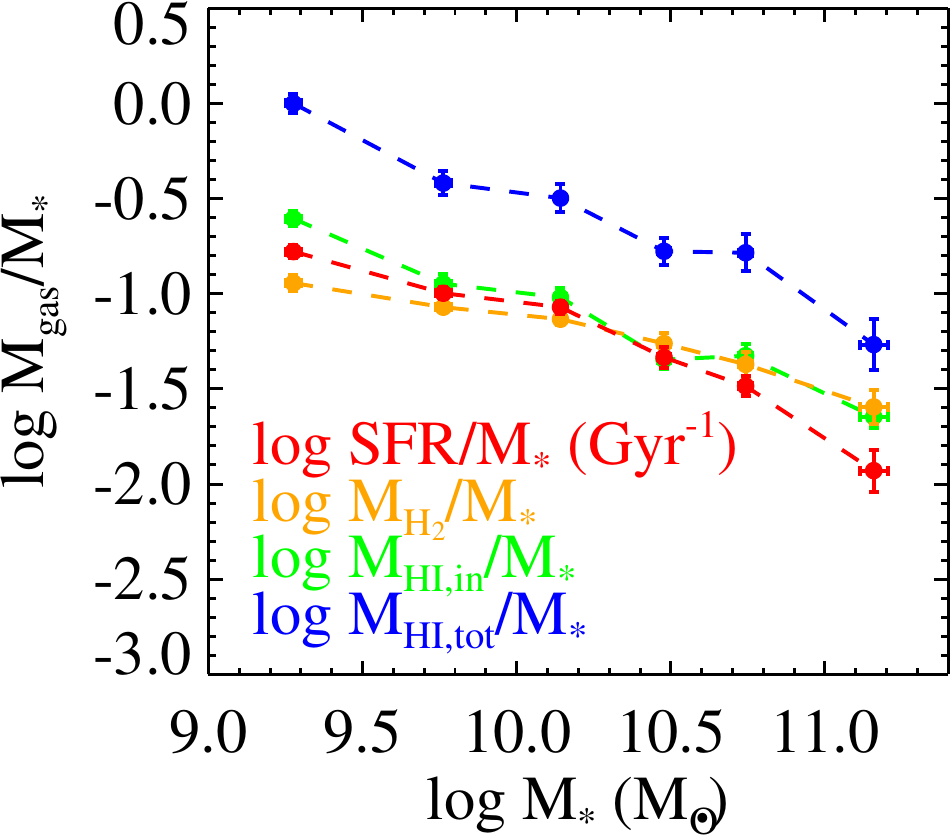}
\includegraphics[width=5.8cm,angle=0]{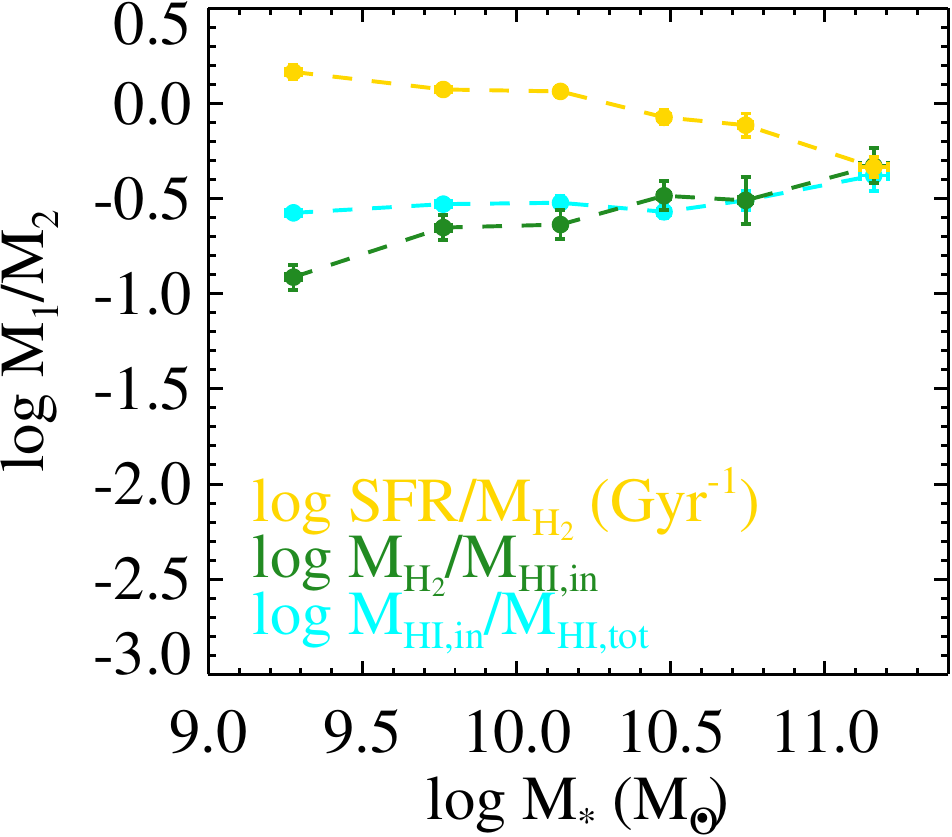}
\includegraphics[width=5.8cm,angle=0]{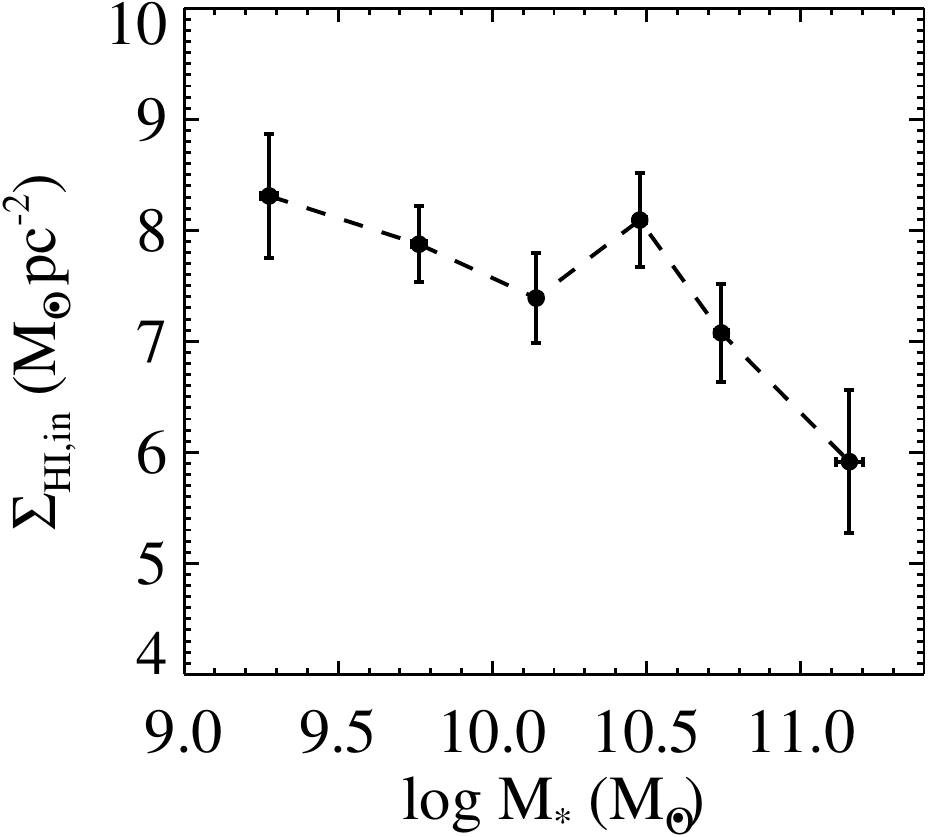}

\caption{Properties of neutral gas along the SFMS for the main sample. Mean values and bootstrapped error bars are calculated $\pm$0.4 dex from the SFMS \citep{Saintonge16} in each $M_*$ bin. Because the main purpose of the figure is to compare the dynamic range of parameters, we have added offsets of 9 and 8.5 dex in the left two panels to shift the $SFR$ related curves close to the other curves.   }
\label{fig:alongSFMS}
\end{figure*}

In Fig.~\ref{fig:alongSFMS} we show how the neutral gas related parameters vary along the SFMS \citep[equation from][]{Saintonge16} as a function of $M_*$, by averaging over disk galaxies (in total 132 galaxies from the main sample) which have their $SFR$ within $\pm$0.4 dex from the SFMS a fixed $M_*$. 

The left panel shows ratios of gas masses and $SFR$ over $M_*$, which reflects the abundance of the gas reservoir in each state along the process of forming young stars. The middle panel shows the ratios of gas masses and $SFR$ over gas masses, which reflects the efficiency of each step of the total $\hi$ reservoir being converted to stars. The y axis of these two panels are displayed with the same width of 3.5 dex, so we can directly compare the slope of the observed trends (i.e. extent of variation of the parameters as a function of $M_*$) in these two panels. We can see that $\fHI$, $\fHIin$, $\fHtwo$ and $sSFR$ show similarly strong decreases, while $\mHIin/\mHI$, $\mHtwo/\mHIin$ and $SFR/\mHtwo$ vary relatively weakly as a function of $M_*$. 

Finally, the right panel shows that $\SHIin$ roughly decreases as a function of $M_*$.

\subsection{Trends as a function of deviation from the SFMS}
 We now study the trend of neutral gas properties when disk galaxies of a given $M_*$ deviate from the SFMS, i.e. as a function of $\Delta \log SFR=\log SFR/SFR_{SFMS}$, where $SFR_{SFMS}$ is the mean SFR of star-forming galaxies on the SFMS at a given $M_*$. 

\begin{figure*}
\centering
\includegraphics[width=7.cm,angle=0]{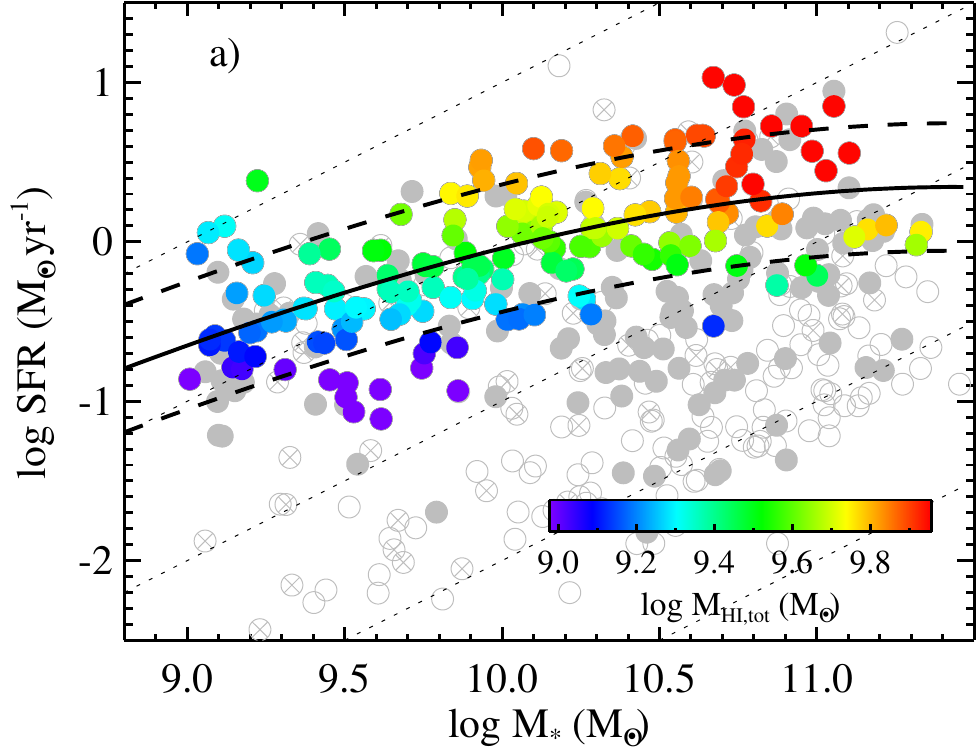} 
\includegraphics[width=7.cm,angle=0]{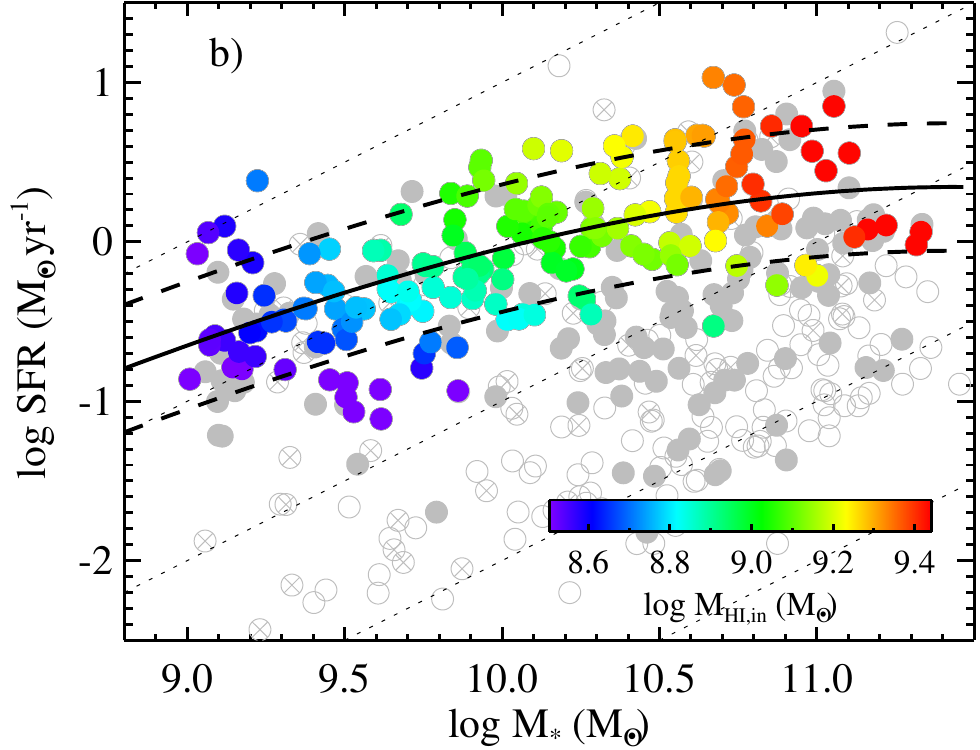} 

\includegraphics[width=7.cm,angle=0]{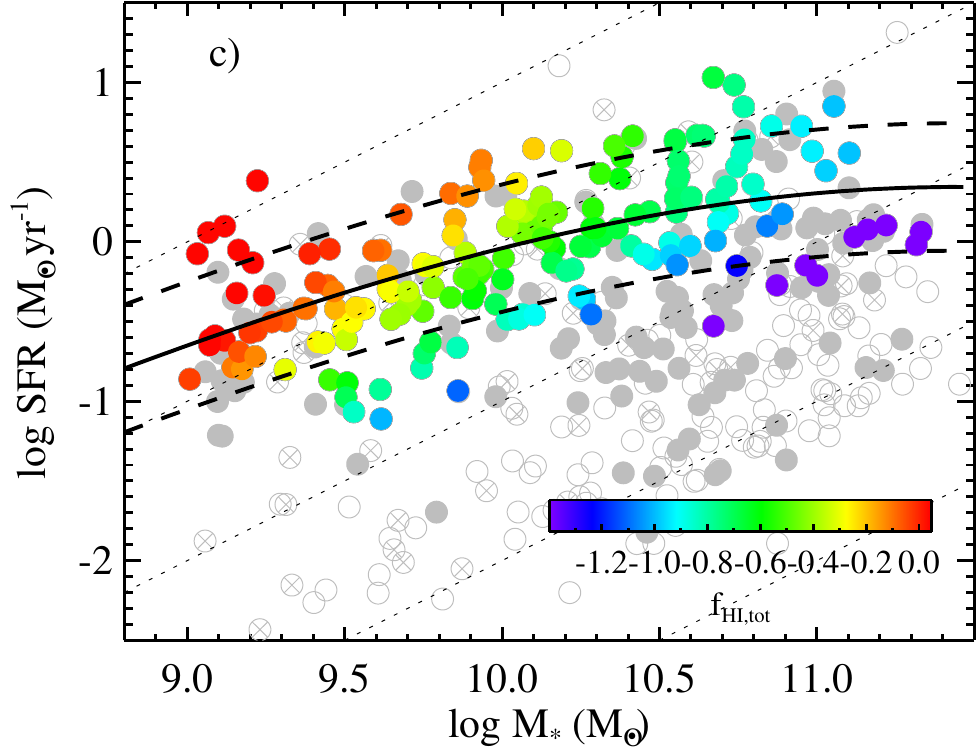} 
\includegraphics[width=7.cm,angle=0]{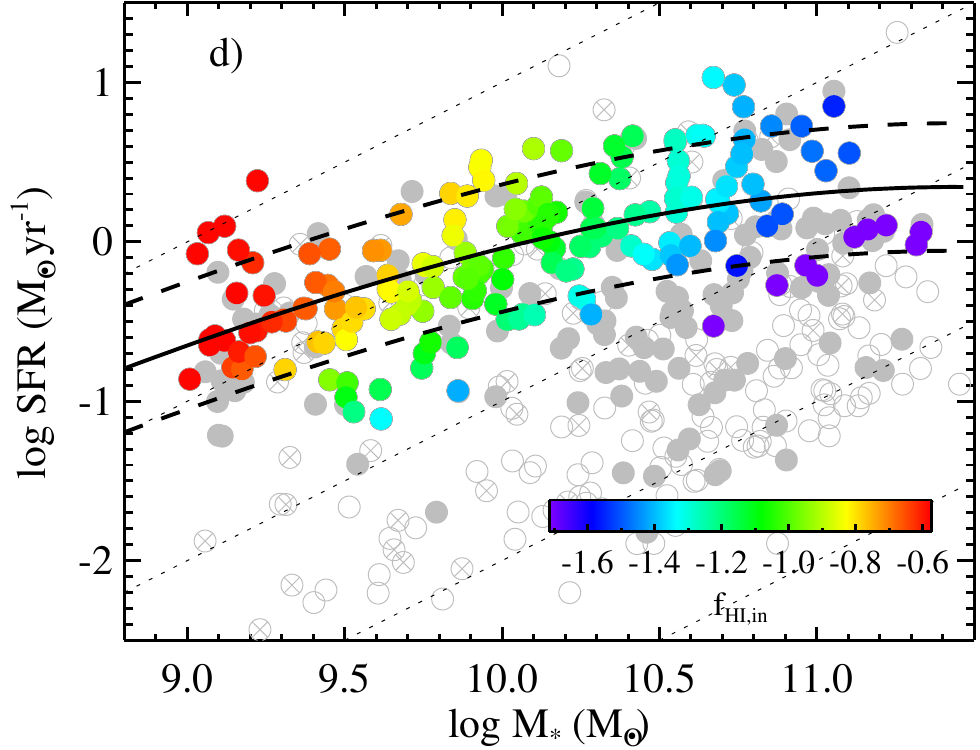} 

\includegraphics[width=7.cm,angle=0]{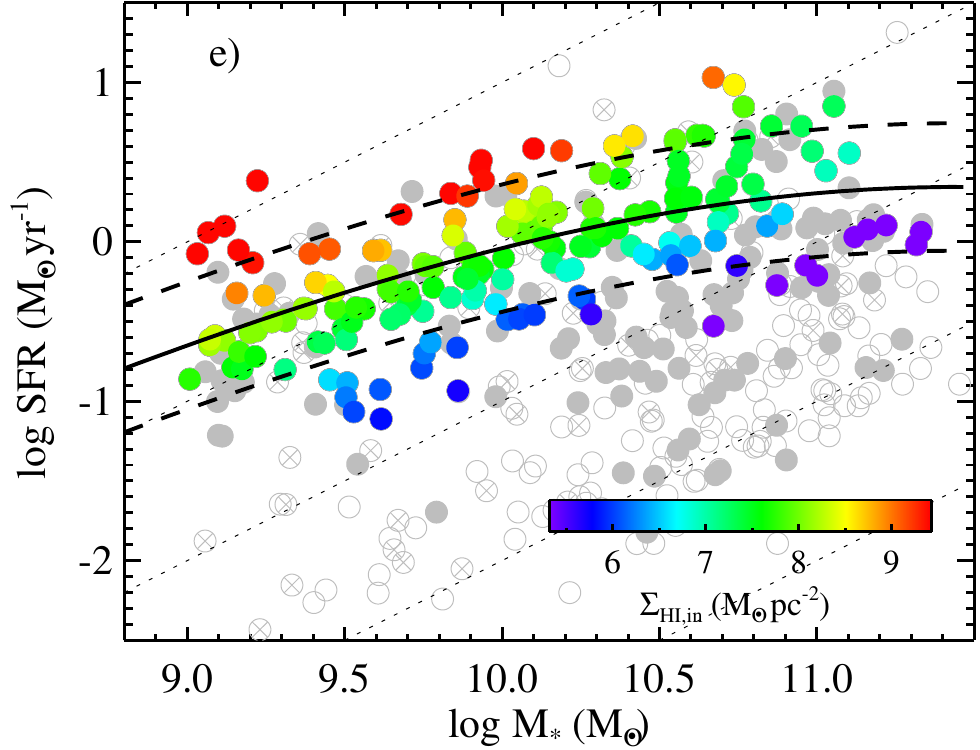} 
\includegraphics[width=7.cm,angle=0]{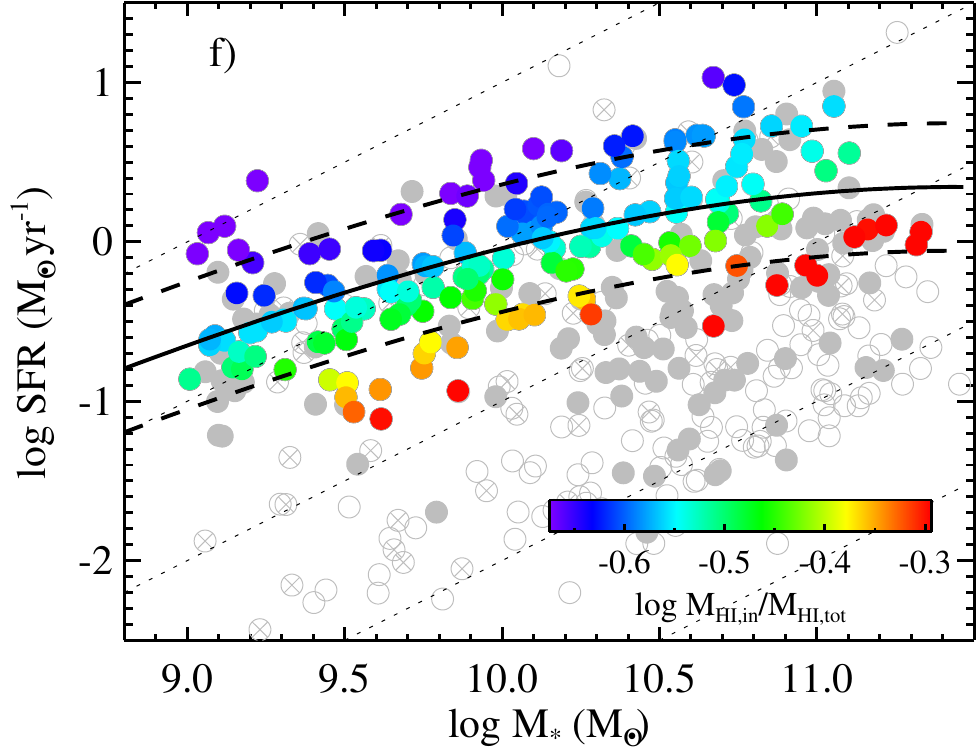} 

\includegraphics[width=7.cm,angle=0]{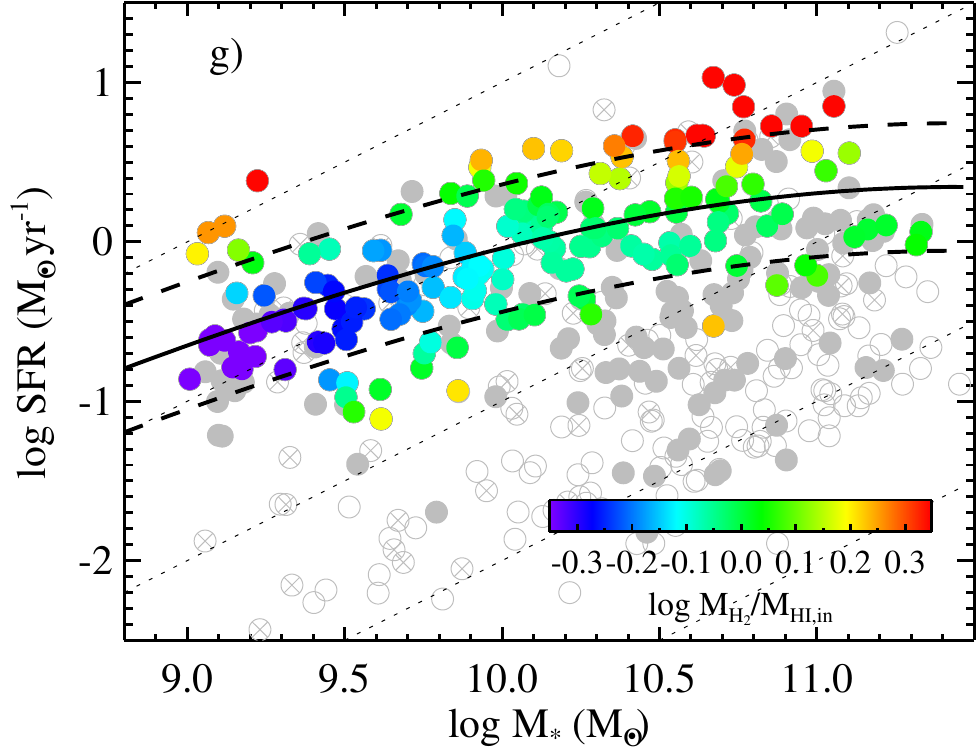} 
\includegraphics[width=7.cm,angle=0]{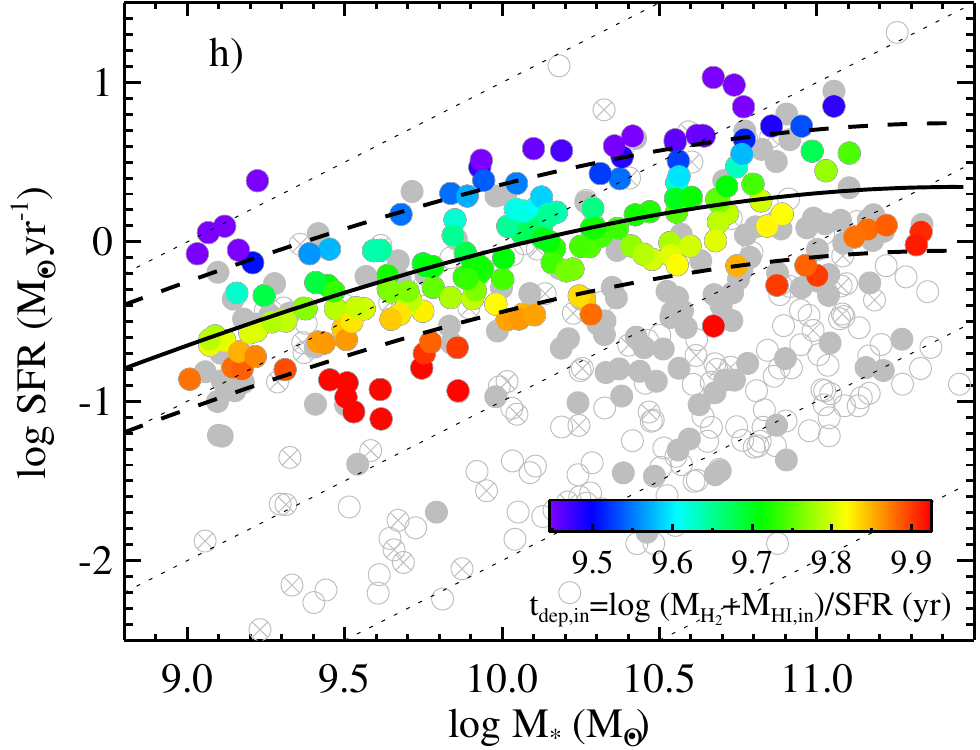} 
\caption{Properties of neutral gas in the space of $SFR$ versus $M_*$. The main sample galaxies are shown as color coded dots, the xCOLD GASS bulge-dominated ($r_{90}/r_{50}>2.7$) galaxies which are detected in $\hi$ and CO are shown as solid grey dots, the xCOLD GASS disk$\slash$bulge-dominated galaxies which are undetected in $\hi$ or CO are shown as open grey circles with$\slash$without crosses. The color coded quantities are LOESS-smoothed \citep{Cappellari13} to highlight the main trend. The color bars highlight the range between 5 and 95 percentiles of the LOESS-smoothed distributions. The solid curve show the mean position of the SFMS, and the dashed curves are vertically $\pm$0.4 dex from solid curve. The diagonal dotted lines have slopes of unity and show positions of constant sSFR, with separations of 1 dex.}
\label{fig:SFMS}
\end{figure*}

The results are presented in Fig~\ref{fig:SFMS}. The most clear trends we see is that, at a fixed $M_*$, higher $SFR$ is on average related with higher $\mHI$ (panel a), higher $\mHIin$ (panel b), higher $\fHI$ (panel c)\footnote{We notice that the slope for lines of constant $\fHI$ in panel a looks smaller than that of $\fHIin$ in panel d, which may give an false impression that for a given stellar mass, $\fHIin$ increases faster with $SFR$ than $\fHI$. We point out that the false impression is due to different color scales of the two panels, and panel d clearly shows the trend of $\mHIin/\mHI$ decreasing as a function of $SFR$.}, higher $\fHIin$ (panel d), higher $\SHIin$ (panel c), lower $\mHIin/\mHI$ (panel d), and lower $t_{\rm dep,in}$ (panel f). Galaxies which have $\Delta \log SFR>0.4$ have the shortest $t_{dep}\lesssim 3 \times 10^9~yr$. 

\begin{figure*}
\centering
\includegraphics[width=14cm,angle=0]{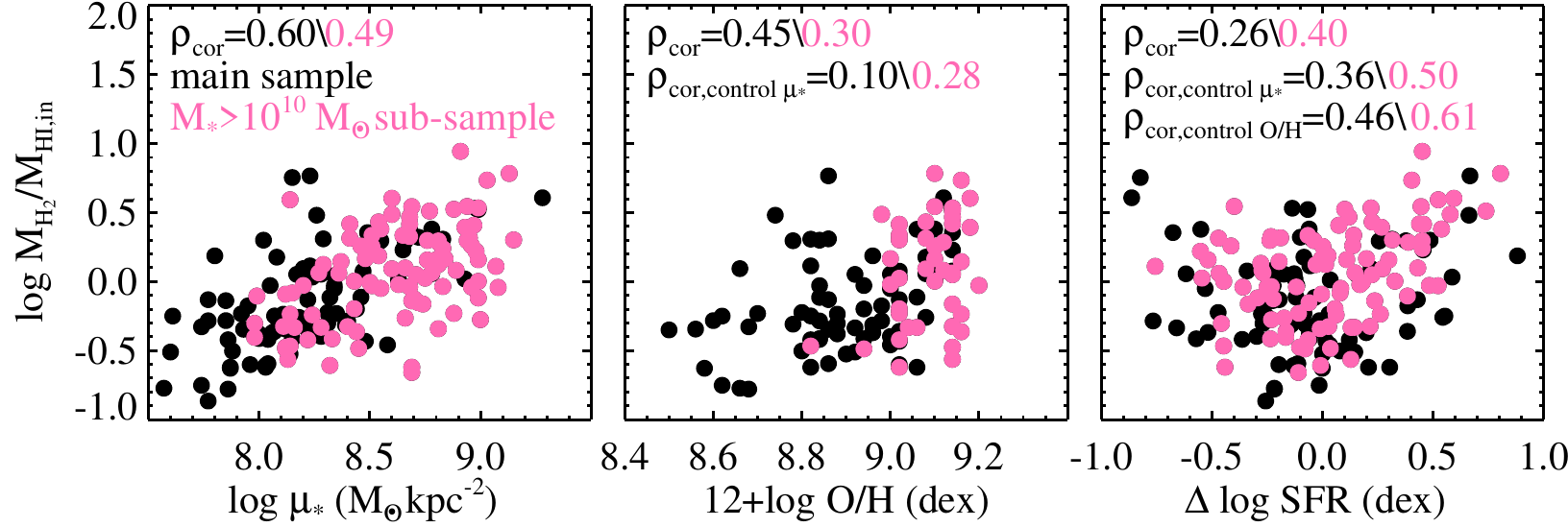} 
\caption{The dependence of $\mHtwo/\mHIin$ on other parameters. The main sample is plotted. The dependences on effective stellar mass surface density ($\mu_*$), gas-phase metallicity ($O/H$, 117 main sample galaxies with reliable measurements from the MPA$\slash$JHU catalog) and $\Delta \log SFR$ are investigated. Pearson correlation coefficients ($\rho_{cor}$) and partial correlation coefficients with the effect of $\mu_*$ or $O/H$ removed ($\rho_{cor,control~\mu_*}$ and $\rho_{cor,control~O/H}$ respecitvely) are calculated for the two parameters on the x- and y-axis of each panel. The data points and correlation coefficients of massive galaxies which have $M_*>10^{10}~M_{\odot}$ are highlighted in pink. }
\label{fig:H2HI_dependence}
\end{figure*}

We also see that galaxies with higher $SFR$ tend to have higher $\mHtwo/\mHIin$ when $M_*>10^{10} M_{\odot}$ (panel e). Such a trend is not observed in low-mass galaxies which have $M_*<10^{10} M_{\odot}$, but it is unclear whether it is affected by selection effects, as the low-$M_*$ and low-$SFR$ galaxies are close to the CO detection limit of xCOLD GASS \citep{Saintonge17}. 
Because $\mHtwo/\mHIin$ is also expected to be correlated with stellar surface density and gas-phase metallicity, one question is whether its enhancement in high-$SFR$ and $M_*>10^{10}~M_{\odot}$ galaxies (panel e of Fig~\ref{fig:SFMS}) is due to a possibly systematic increase in stellar surface density or metallicity. Fig.~\ref{fig:H2HI_dependence} confirms the strong and weak correlation of $\mHtwo/\mHIin$ with $\mu_*$ and $O/H$, respectively. Calculation of the partial correlation coefficients \footnote{In order to calculate the partial correlation coefficient between parameters $A$ and $B$ with parameter $C$ controlled, one firstly derive the best linear fits of $A$ versus $C$ and $B$ versus $C$, and then the Pearson correlation coefficient between the offsets of both relations is calculated as the final partial correlation coefficient.} suggests that when $M_*>10^{10}~M_{\odot}$, the trend of $\mHtwo/\mHIin$ increasing with $\Delta \log SFR$ becomes stronger when the effect of $\mu_*$ or gas-phase metallicity $O/H$ is removed. 

There might be a concern that the anti-correlation between $t_{\rm dep,in}$ and $SFR$ is due to $SFR$ being present in both axes. The same applies to the relation between $\fHI$ ($\fHIin$) and $M_*$. Although this is true, the fact that the relations are not exactly linear shows that at least part of the correlation is not induced by plotting repeated quantities. Indeed, the relation between $t_{\rm dep,in}$ and $SFR$ has a negative slope, which indicates that SFR does increase linearly as a function of $M_{\rm gas}$, but in a faster (super-linear) way. Similarly, the anti-correlation between $\fHI$ ($\fHIin$) and $M_*$ suggests that $\mHI$ ($\mHIin$) increases as a function of $M_*$ in a sub-linear way.

Another possible concern is how random errors on the $SFR$ and stellar masses affect these relations. To address this, we calculate the error-corrected Pearson
correlation coefficients for these relations, accounting for the estimated errors of $SFR$ (typically 0.1 dex) and $M_*$ (typically 0.09 dex). We use the equation of \citet{Charles05}:
\begin{equation}
\begin{aligned}
r_{Tx,Ty}=(r_{x,y}- r_{Ex,Ey}*\sqrt{e_{xx}*e_{yy}} - r_{Tx,Ey}*\sqrt{r_{xx}*e_{yy}} \\
 - r_{Ex,Ty}*\sqrt{e_{xx}*r_{yy}}  )/\sqrt{r_{xx}*r_{yy}} ,
 \end{aligned}
\end{equation}
where $x$ and $y$ are measurements of two parameters, $Ex$ and $Ey$ are their errors, and $Tx$ and $Ty$ are their true values, so that $x=Tx+Ex$ and $y=Ty+Ey$. $r_{i,j}$ is Pearson correlation coefficient between two quantities $i$ and $j$, $r_{ii}$ is the reliability of a measurement $i$, and $e_{ii}$ is the proportion of variance in measurement $i$ that is due to error. So $r_{x,y}$ is the directly calculated Pearson correlation coefficient of the measurement, the correction terms with $r_{Ex,Ey}$, $r_{Tx,Ey}$ and $r_{Ex,Ty}$ correct for contributions from errors being correlated with each other or with the measurements, and the denominator of the equation corrects for attenuation of the intrinsic correlation coefficients ($r_{Tx,Ty}$) due to unreliable measurements.
The error of each measurement ($Ex$ and $Ey$) is simulated as a random value taken from a normal distribution with $\sigma$ equivalent to the measurement uncertainty ($\sigma_x$ and $\sigma_y$). We further use the measurements x and y to approximate the intrinsic values $Tx$ and $Ty$ when calculating $r_{Tx,Ey}$ and $r_{Ty,Ex}$. We derive a corrected Pearson correlation coefficient of -0.52 for the relation between $t_{dep,in}$ and $SFR$,  -0.63 for the relation between $\fHI$ and $M_*$, and -0.73 for the relation between $\fHIin$ and $M_*$. These coefficients indicate strong correlations after taking into account the correlation of parameters with measurement errors. The approximation of approximate $Tx$ ($Ty$) with $x$ ($y$) while calculating $r_{Tx,Ey}$ ($r_{Ty,Ex}$) tends to over-estimate the relevant correction terms and hence under-estimate $r_{Tx,Ty}$, because the addition of $Ex$ ($Ey$) with respect to $Tx$ ($Ty$) is correlated with $Ey$ ($Ex$) for the parameter pairs considered here. So the estimated error-corrected Pearson correlation coefficients $r_{Tx,Ty}$ are conservative and the correlation of these three parameter pairs are truly strong. 

To summarize, the results suggest that SFR enhanced galaxies tend to have built a large and dense $\hi$ reservoir within the stellar disks, and achieved efficient atomic-to-molecular gas conversion in at least the $M_*>10^{10}~M_{\odot}$ galaxies; they are likely to deplete the neutral gas on the stellar disks quickly, but  gas inflows seem not very efficient (low $\mHIin/\mHI$) and thereby have built very extended $\hi$ disks (instead of concentrating the gas into the center and trigger starbursts) in these galaxies.

\subsection{Additional dependence on relative stellar compactness}
The average stellar surface density in the central 1-kpc region, $\Sigma_{*,1}$ quantifies the central stellar compactness \citep{Cheung12, Fang13, Woo15, Tacchella16b, Mosleh17, Whitaker17, Wang18}. Because the absolute compactness is correlated with $M_*$, it is useful to derived the mean relation between $M_*$ and $\Sigma_{*,1}$, and calculate $\Delta \Sigma_{*,1}$, the deviation of $\Sigma_{*,1}$ from the mean relation at a given $M_*$. $\Delta \Sigma_{*,1}$ serves as an indicator of relative compactness at a given $M_*$.  The relation between $\Delta \Sigma_{*,1}$ and $\Delta_{SFR}$ has become a useful tool in investigations about the compaction scenario \citep{Barro17, Whitaker17, Wang18, Luo19}. Galaxies are observed to distribute in an ``L'' shape in the space of $\Delta \Sigma_{*,1}$ and $\Delta_{SFR}$, where high $\Delta_{SFR}$  galaxies have a wide range of $\Delta \Sigma_{*,1}$, but quenched galaxies almost all have high $\Delta \Sigma_{*,1}$. The ``L'' shape is consistent with the prediction of the compaction model, where galaxies first develop a compact stellar center with efficient star formation, before they cease the star formation \citep{Dekel14}. 

\begin{figure*}
\centering
\includegraphics[width=8.8cm,angle=0]{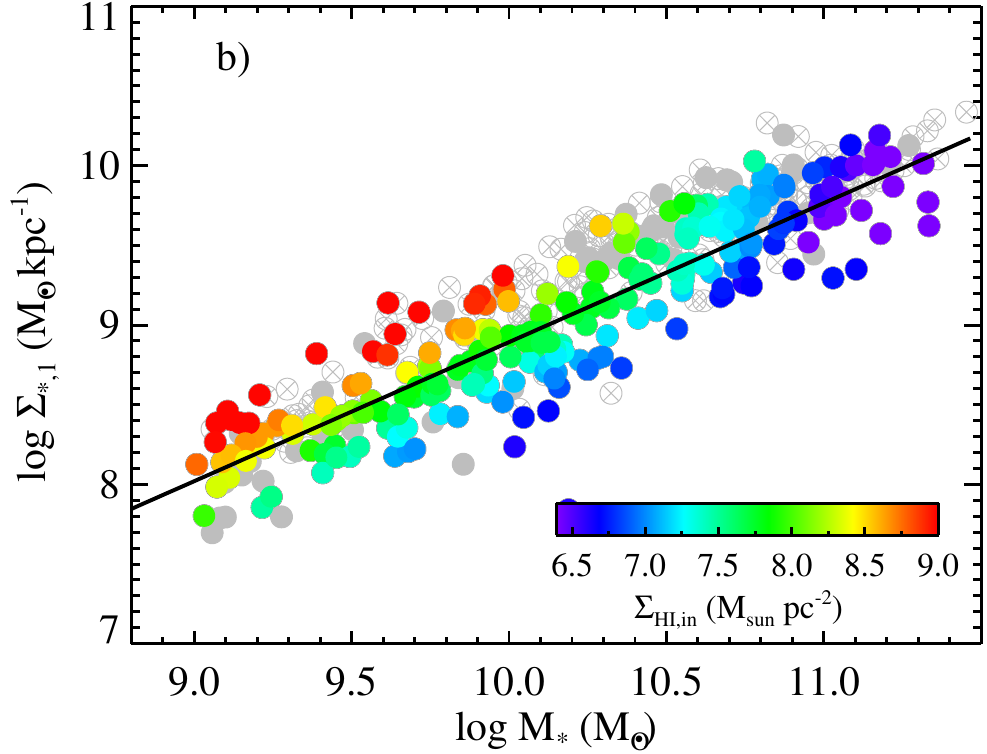}
\includegraphics[width=8.8cm,angle=0]{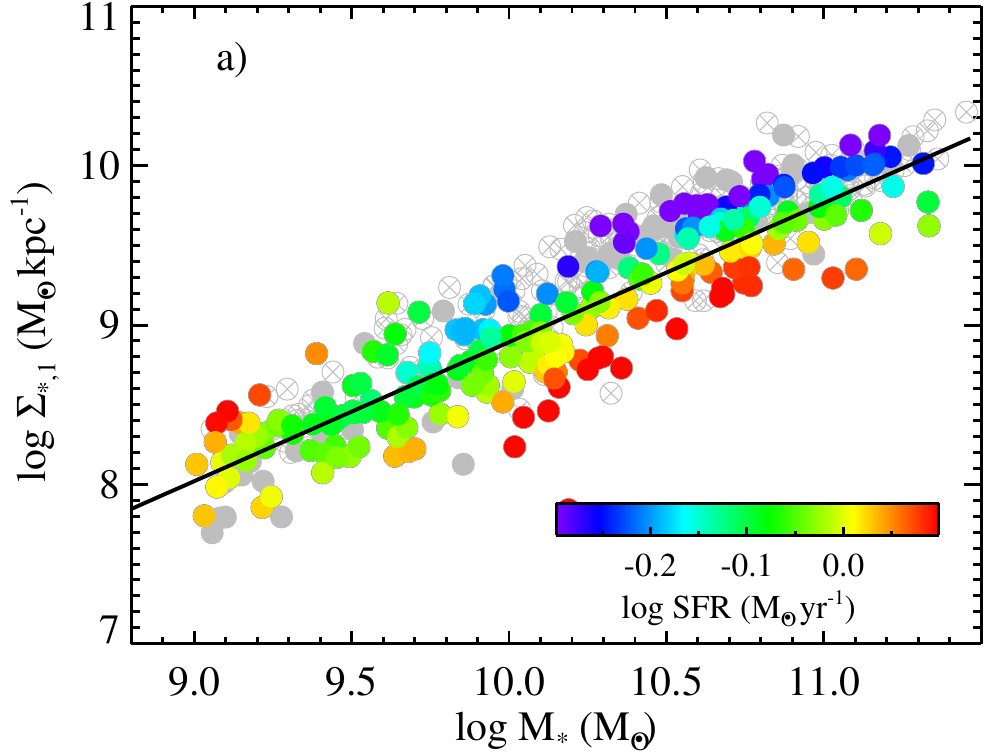}

\includegraphics[width=8.8cm,angle=0]{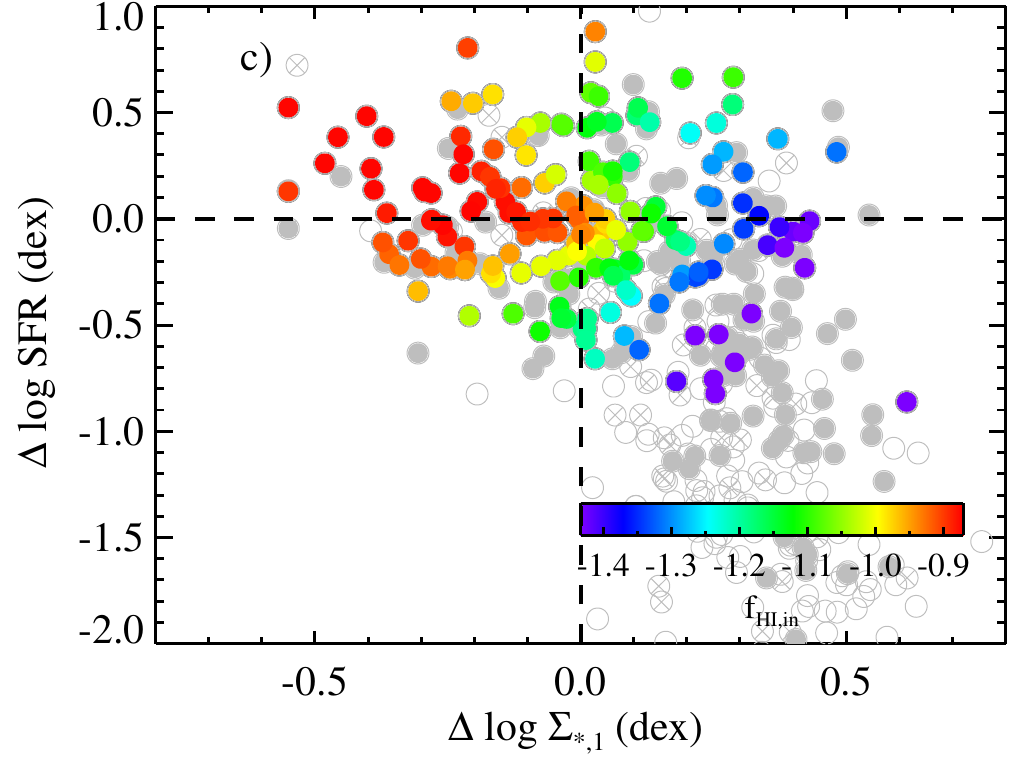}
\includegraphics[width=8.8cm,angle=0]{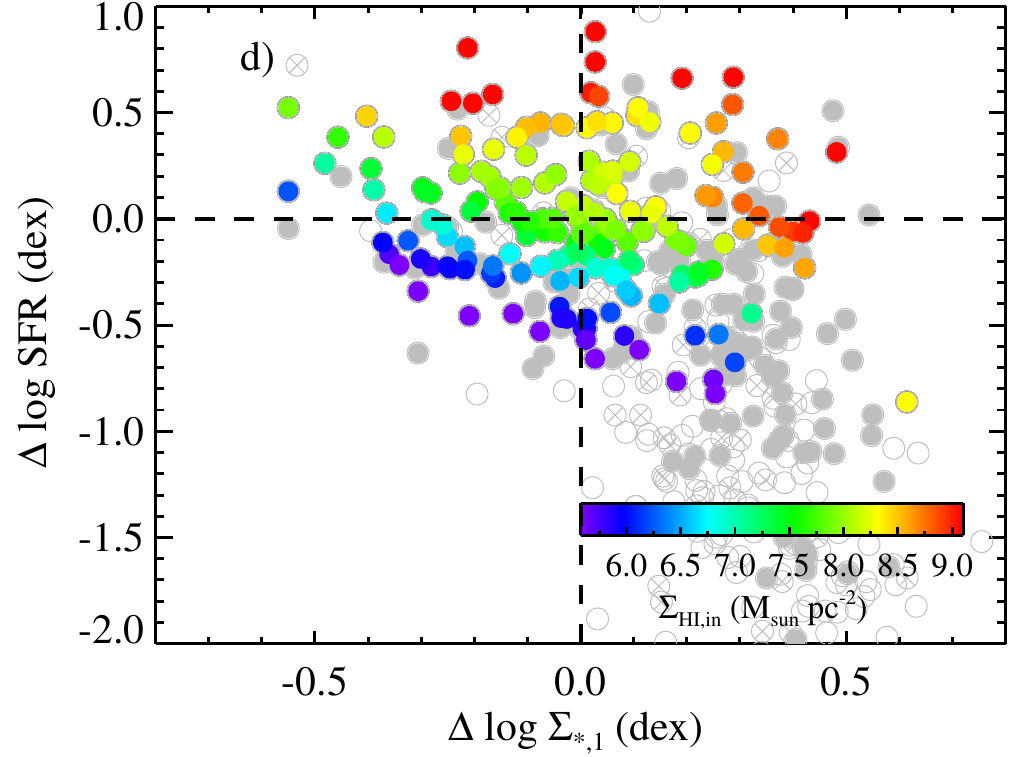}
\includegraphics[width=8.8cm,angle=0]{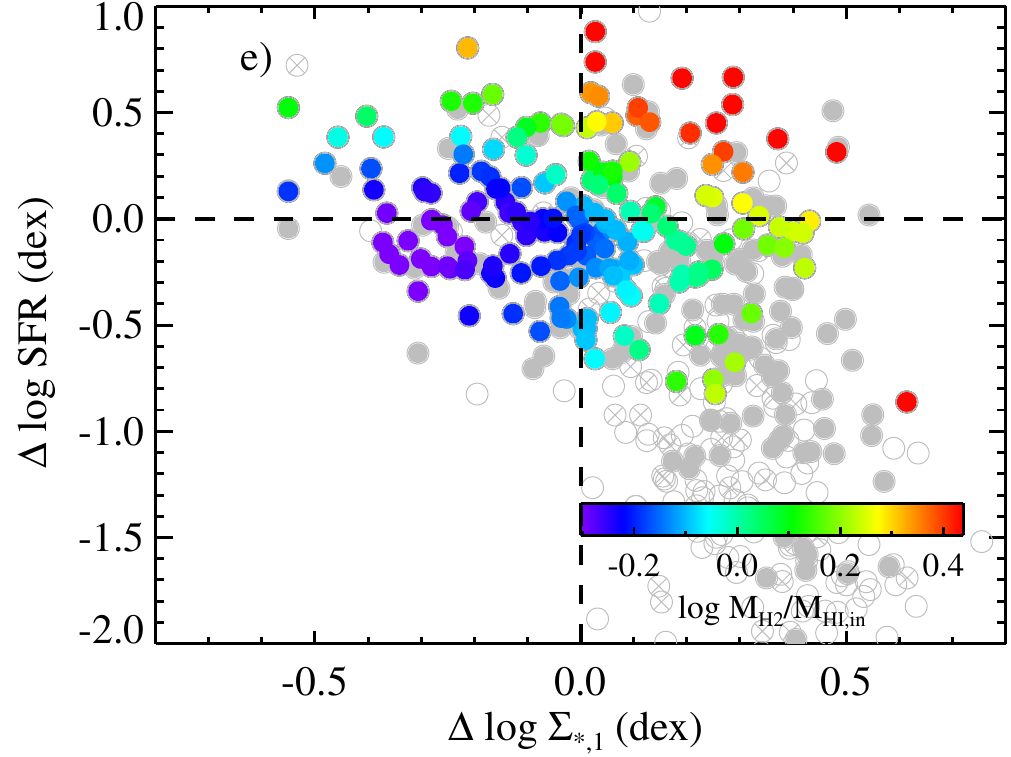}
\includegraphics[width=8.8cm,angle=0]{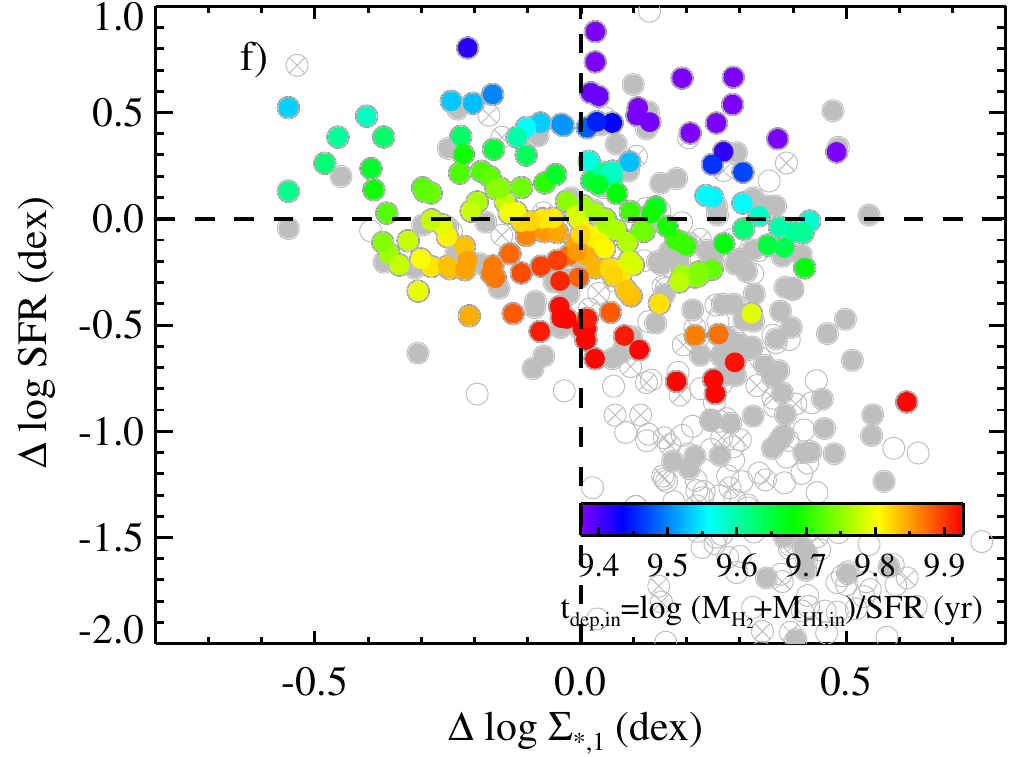}
\vspace{0.2cm}
\caption{The $\Sigma_{*,1}$ versus $M_*$ relation, and the distribution of $\mHIin$ related parameters in the space of $\Delta log SFR$ versus $\Delta log \Sigma_1$. The main sample galaxies are shown as color coded dots, the xCOLD GASS bulge-dominated ($r_{90}/r_{50}>2.7$) galaxies which are detected in $\hi$ and CO are shown as grey dots, and the xCOLD GASS disk$\slash$bulge-dominated galaxies which are undetected in $\hi$ or CO are shown as open grey circles with$\slash$without crosses. The color coded quantities are LOESS-smoothed \citep{Cappellari13} to highlight the main trend. }
\label{fig:dsfdsz}
\end{figure*}

Panels a and b of Fig.~\ref{fig:dsfdsz} show how $SFR$ and $\SHIin$ are distributed in the space of $\Sigma_{*,1}$ versus $M_*$ respectively. At a given $M_*$, more compact galaxies tend to have lower $SFR$ when $M_*>10^{10} M_{\odot}$;  more compact galaxies also tend to have higher $\SHIin$. The trend of decreasing SFR with increasing stellar compactness (at a given $M_*$) is consistent with \citet{Saintonge16}.

We then look into gas properties in the space of $\Delta \Sigma_{*,1}$ versus $\Delta_{SFR}$, which can be more conveniently compared to the compaction model. From panel c to f, it is clear that our sample is selected against the really passive galaxies which typically have high $\Sigma_{*,1}$ (the passive extension of the ``L'' shape is grey in the figure). Our study hence focuses on the fueling (toward compaction) but misses the quenching (following compaction) part of the compaction model \citep{Tacchella15}.  

Panel c of Fig.~\ref{fig:dsfdsz} shows that less compact galaxies have higher $\fHIin$ than more compact galaxies. At low compactness, there is no significant correlation between $\fHIin$ and $\Delta \log SFR$. The lowest $\fHIin$ is found in the galaxies with high compactness and low SFR-enhancement. 

 $\SHIin$ increases (panel d) and $t_{dep,in}$ decrease (panel e) in the direction of $\Delta \log SFR$ and $\Delta \Sigma_{*,1}$. $\mHtwo/\mHIin$ does not show a similarly strong trend as $\SHIin$ or $t_{dep,in}$, but the highest values tend to be found where $\Delta \log SFR>0$ and $\Delta \Sigma_{*,1}>0$ (panel c). Hence those compact, SFR enhanced galaxies tend to build a denser neutral gas reservoir but deplete it more efficiently than less compact galaxies.  

\section{Discussion}
We have developed a new method to estimate the $\hi$ mass ($\mHIin$) and surface densities ($\SHIin$) within the optical $r_{90}$ for disk galaxies. $\mHIin$ serves as an intermediate state between the total $\hi$ reservoir and the $\htwo$ disk. Although a large fraction of $\hi$ in $\hi$-rich galaxies lies beyond the stellar disk, $\hi$ within the stellar disk (compared to $\htwo$) is still likely the dominant reservoir for star formation (panel e of Fig.~\ref{fig:SFMS}), making $t_{dep,in}$ ($>3$ Gyr) derived in this paper considerably longer than the $\htwo$ depletion time \citep{Saintonge17}. We discuss below how the newly derived $\mHIin$ properties confirm or alter our previous understanding of galaxy evolution around the SFMS based on observations of global $\hi$ measurements. {\bf We emphasize that all results and discussions are limited to late-type disk galaxies.} There are 348 and 142 galaxies from the xGASS-disk and main samples respectively, which are identified as the central galaxy of groups (including isolated galaxies) in the catalog of \citet{Yang07}. We note that all the $\hi$-related trends presented in the paper do not change if we limit the analysis to the late-type central galaxies. 

\subsection{Trends revealed with the estimate of $\SHIin$}
\label{sec:discuss1}
Because the real physical relation is between the surface densities of the SFR and the neutral gas, $\SHIin$ is a better tracer of SFR enhancement than $\mHI$ or $\mHIin$ (panel c and d of Fig.~\ref{fig:dsfdsz}). The trend is about building a dense gas reservoir to onset and sustain highly efficient star formation, and was not observationally demonstrated in the context of galaxy evolution for a $M_*$-selected and statistically significant sample of late-type galaxies. 

We emphasize that the link between SFR and the inner $\hi$  appears similar to but different from the Kennicutt-Schmidt law of star formation \citep{Kennicutt98}, because $\hi$ serves not as the direct material (compared to the molecular gas) for star formation, but an intermediate state in the star formation fueling process. At an earlier stage of this process, an excess of warm ($10^4$ K) ionized gas in the CGM is observed around star-forming galaxies with respect to passive galaxies in the local universe \citep{Borthakur15, Borthakur16}. It is also well accepted that star-forming galaxies tend to be globally $\hi$-rich \citep{Catinella10,Catinella18}. Our study further resolves the fueling process and find that the $\hi$ within the stellar disks also builds up when galaxies tend to have high $SFR$. 
This trend is not so obvious before we quantify it, because $\hi$ is not the direct material for forming stars \citep{Bigiel08, Wang17}. It would be possible for one to speculate that at a given $M_*$ the galaxies with the highest SFR could have the lowest $\mHIin$, if $\hi$ in these galaxies were converted to $\htwo$ as efficiently as in the starbursting galaxies at high redshift \citep{Tacconi18}. Our result suggests that this speculations is not true.

More compact galaxies on average may need higher inner $\hi$ surface densities to achieve the same extent of SFR enhancement (higher SFR surface densities due to smaller disk sizes) than less compact galaxies (panel d of Fig.~\ref{fig:dsfdsz}). These compact, star formation enhanced galaxies also deplete their neutral gas within the stellar disks more quickly than other galaxies (panel c and d of Fig.~\ref{fig:dsfdsz}). Their existence implies a possible evolutionary path for disk galaxies to develop a central bulge and cease star formation simultaneously, if gas replenishment is suppressed (e.g. if gas accretion is suppressed by a massive halo, \citealp{Rees77, Birnboim03, Keres05, vandeVoort11, Gabor15}, or lack of gas inflow due to weak disk instabilities, \citealp{Noguchi98, Bournaud07, Dekel09, Cacciato12}). This is exactly the way the compaction model predicts how galaxies cease their star formation \citep{Dekel14, Zolotov15, Tacchella15}. These compact and $SFR$-enhanced galaxies are likely precursors of such an evolution (also see \citealt{Ellison18}).

\begin{figure}
\centering
\includegraphics[width=7.5cm,angle=0]{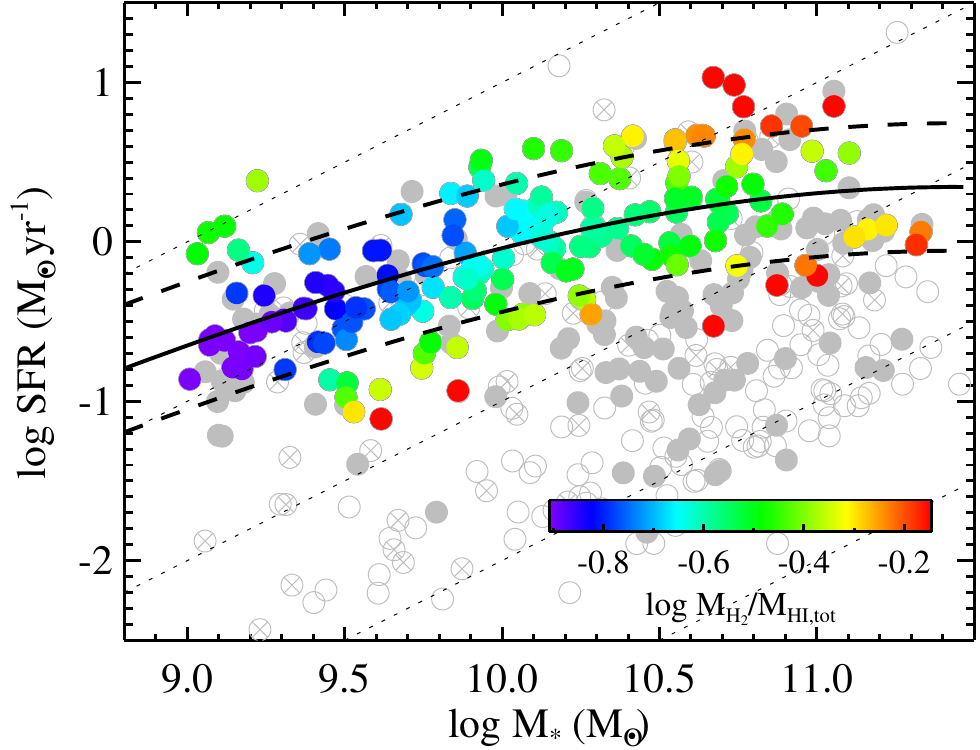}
\vspace{0.2cm}
\caption{Distribution of $\mHtwo/\mHI$ in the space of $SFR$ versus $M_*$. Similar to Fig.~\ref{fig:SFMS}, but the dots are colour coded by $\mHtwo/\mHI$.}
\label{fig:SFMS2}
\end{figure}

The atomic-to-molecular conversion ($\mHtwo/\mHIin$) contributes to boosting the SFR, at least when $M_*>10^{10}~M_{\odot}$ (panel g of Fig.~\ref{fig:SFMS} and right panel of \ref{fig:H2HI_dependence}). This effect was not observed with global measurements \citep{Catinella18}, but is predicted in the compaction model \citep{Dekel14, Zolotov15, Tacchella15}. We can also see in Fig.~\ref{fig:SFMS2} that the global measurement $\mHtwo/\mHI$ does not increase monotonically with $SFR$ at a given $M_*$, but reverse to show an increase when the galaxies are below the SFMS. As a result the Pearson correlation coefficient between $\mHtwo/\mHI$ and $\Delta \log SFR$ is only 0.14 (0.46 for the $\mHtwo/\mHIin$ versus $\Delta \log SFR$ relation), and the partial correlation coefficient with the effect of $\mu_*$ controlled is 0.28 (0.61 for the $\mHtwo/\mHIin$ versus $\Delta \log SFR$ relation). This reversal in trend is likely due to the shrink of the $\hi$ disks instead of enhanced conversion efficiency at low $SFR$.

Finally, a relatively low efficiency of gas inflows (lower $\mHIin/\mHI$ for more star-forming galaxies) may be a major obstacle in fueling the SFR (panel f of Fig.~\ref{fig:SFMS}). Theoretically, under a CDM cosmological context, the large-scale accreting gas has a high specific angular moment \citep{Mo98}, while the star formation induced accretion of gas from the inner parts of CGM (fountain gas) also tends to have high specific angular moment due to a mixing with the high-angular momentum CGM \citep{Grand19}, and possibly also due to the suppressing of low-angular momentum fountains with short dynamic times \citep{Marasco12}. The gas with a high specific angular momentum tends to build an extended $\hi$ disk. 
We would expect much more vigorous star formation than observed, if the massive $\hi$ in the outer disks of $\hi$-rich galaxies could be efficiently driven to the center. The relatively low efficiency might be related to the fact that disk instabilities and tidally interacting frequencies are relatively low at low redshift compared to the high redshift \citep{Noguchi98, Bournaud07, Dekel09, Cacciato12}. It may also be related to the theoretically predicted self-regulation of low-redshift disks \citep{Krumholz18}, where inflows are driven by disk instabilities (associated with low gas velocity dispersion and low Toomre $Q$), and a high inflow rate will result in high $SFR$ and increased gas velocity dispersion (due to stellar feedbacks and gravitational heating of the inflow gas) which then suppress the disk instabilities (high Toomre $Q$). Such a mechanism prevents strong inflows and maintains the extended $\hi$ disks in $\hi$-rich galaxies. This feature was not described in classical compaction models \citep[e.g.][]{Dekel14, Zolotov15, Tacchella15}, and may serve as a new constraint for these models in the local universe.

\subsection{Known trends of $\mHI$ confirmed by $\mHIin$}
One major feature of the SFMS is its slope being shallower than one. It was found that along the SFMS $\fHI$ and $\fHtwo$ drops much faster than $\mHtwo/\mHI$ as a function of $M_*$ \citep{Saintonge16, Saintonge17, Catinella18}. Hence it is the shrinking of the gas reservoir, rather than a bottleneck in converting the atomic gas to the molecular gas, that plays a major role in the flattening of the SFMS. Our results (both $\mHIin/\mHI$ and $\mHtwo/\mHIin$ vary little with $M_*$ along the SFMS, middle panel of Fig.~\ref{fig:alongSFMS}) elaborate that when the SFMS flattens, the major bottleneck (of forming stars) is neither in driving $\hi$ inward to the stellar disks, nor in converting $\mHIin$ to $\mHtwo$ with the stellar disks. $SFR/M_*$, $\mHtwo$ and $\mHIin/M_*$ drop similarly fast as $\mHI/M_*$ along the SFMS (left panel of Fig.~\ref{fig:alongSFMS}), hence it is indeed likely (as concluded by \citealt{Saintonge16}) that the global $\hi$ abundance (as the first step of fueling from $\hi$ to star formation) strongly regulates the slope of the SFMS.
 
The most star-forming galaxies rarely go far ($>0.4$ dex in SFR) above the SFMS. A short depletion time of the total neutral gas and molecular gas was found for the star-bursting galaxies with respect to normal star-forming galaxies \citep{Genzel15, Silverman15, Scoville16, Saintonge17}. The trend is observationally related to the super-linear nature of the Kennicutt-Schimid law of star formation \citep{Kennicutt98}. The more efficient depletion is taken as the major confinement for the upper envelope of the $SFR$-$M_*$ relation in the compaction model \citep{Tacchella15}. We show that the trend is still true for disk galaxies when only the neutral gas with the stellar disks is considered (panel h of Fig.~\ref{fig:SFMS}). 

Most of the star-forming galaxies have a scatter of $\pm0.4$ dex in SFR around the SFMS. Previous observations found that $\Delta \log SFR$ at a given stellar mass is strongly set by $\mHI$ and $\mHtwo$ \citep{Whitaker12, Tacchella15, Saintonge16, Saintonge17, Catinella18}. The determining role of the cold gas reservoir on the star-forming status of galaxies was predicted by the compaction model \citep{Tacchella15}. Our results confirm that on average the SFR has to increase with the reservoir of fueling material, globally as well as within the stellar disks for disk galaxies (panels a, b, c, d of Fig.~\ref{fig:SFMS}).

\subsection{Caveats and future perspective}
The main sample (on which most results are based) is very strongly biased toward the SFMS, as shown by the open circles with crosses in Fig.~\ref{fig:SFMS} and \ref{fig:dsfdsz}. Therefore, our results strongly support the compaction scenario (see discussion in Sec.~\ref{sec:discuss1}), but do not directly link to the model predicted consequence that galaxies cease their star formation as a result of compaction \citep{Dekel14, Zolotov15, Tacchella15}. It is unclear which factors or mechanisms most strongly suppress SFR in the passive and massive disk galaxies, i.e. whether $\fHIin$ and $\SHIin$ continuously drops (following the trends in panel acand d of Fig.~\ref{fig:dsfdsz}), or $\hi$ for a given gas density forms too little molecular gas.

We need keep in mind that predicted $\hi$ properties have deviations from the real ones. We caution the reader for unknown systematic dependences of the deviations, because VS used to test our method is a relatively small sample with a complex selection. Processes that break the equilibrium state of galaxies, like tidal effects and episodic gas accretion, may affect the radial distribution of $\hi$. Hence our discussion is limited to the average trend of disk-dominated galaxies where these effects are assumed to be normal. Moreover, $\mHIin$ and $\SHIin$ are estimated within the $r$-band $r_{90}$, which should enclose most but not exactly 90\% of $SFR$ and $\htwo$ gas. This mis-match adds uncertainty to the interpretation of our results, which may have a systematic dependence on the bulge-to-disk ratio of galaxies. We test the significance of this effect by replacing $r_{90}$ with $R_{25}$, the semi-major axis of the 25 mag arcsec$^{-2}$ isophotes, which should be less dependent on the bulge prominence than $r_{90}$. We find that all our trends remain (examples in appendix 2), though with a smaller amplitude due to the averaging of $\SHI$ within a larger aperture than $r_{90}$. We hence conclude that the possible dependence of $r_{90}$ on bulge prominence does not significantly affect our major results and conclusion. 
Nevertheless, confirmation of our results will be needed in the future with real, spatially resolved $\hi$ data, which will be available when the new radio interferometric instruments finish their planned, large surveys of $\hi$ in nearby galaxies in the near future \citep[ASKAP-WALLABY, Apertif, etc.][]{deBlok15, Staveley-Smith15}.

The estimated $\mHIin$ has promising applications in the moderate-redshift SKA and pathfinder HI surveys \citep[DINGO, LADUMA, etc.][]{Meyer09, Holwerda12}, as well as the low-redshift ones \citep[WALLABY]{Staveley-Smith15}, where $\hi$ in most of the galaxies will be unresolved. In addition to serving as an intermediate reservoir for star formation, it has the potential of improving $\htwo$ indicators. Due to the lack of large-sample millimeter surveys, $\htwo$ is often indicated by measurements of dust from infrared photometry or optical spectroscopy data \citep{Brinchmann13, Berta16, Yesuf19}. Because dusts seem to more closely trace the total neutral gas than $\htwo$ \citep{Groves15,Janowiecki18}, and the relations between dusts and neutral gas are different within and beyond the stellar disks (possible due to the different gas-phase metallicities, \citealt{Moran12, Janowiecki18}), the estimated $\mHIin$ provide useful constraints on $\mHtwo$ estimators based on dust properties. Such a potential application will be investigated in a future paper. 

\section{Conclusions}
The method presented offers a useful way to get more information out of global $\hi$ profiles for late-type (disk-dominated) disk galaxies. This will be important for deep interferometric surveys such as DINGO \citep{Meyer09} and LADUMA \citep{Holwerda12}, and also single-dish FAST HI surveys \citep{Li18, Zhang19} as most detections will be unresolved. The method is able to characterise the average $\hi$ surface density within the stellar disk of late-type galaxies, where the gas is directly fueling the star formation. Exploring $\hi$ related parameters in relation to the SFMS, we found that the spread along the SFMS is best characterised by this inner $\hi$ surface density (i.e. this parameter is the most discriminatory perpendicular to the SFMS, especially when considering the conversion of the $\hi$ to the molecular gas or fixing the central compactness of galaxies) among the $\hi$-related parameters. So for studying the spread of late-type galaxies in the SFMS, this is a quantity that one should focus on. 

The trends found are generally consistent with the compaction model of galaxy evolution regulated by the balance between cold gas fueling and star formation depletion \citep{Dekel14, Zolotov15, Tacchella15}.

\vspace{1cm}
We gratefully thank Thijs van der Hulst for useful discussions. This work was supported by the National Science Foundation of China (11721303, 11991052) and the National Key R\&D Program of China (2016YFA0400702). Z. P. acknowledges the support from National Natural Science Foundation of China (NSFC, grant no 11703092).  Parts of this research were supported by the Australian Research Council Centre of Excellence for All Sky Astrophysics in 3 Dimensions (ASTRO 3D), through project number CE170100013. JW further thank support from the ASTRO 3D Science Visitor program at the ICRAR node. 

\bibliographystyle{apj}
\bibliography{HI_in}

\appendix

\section{1. Comparing our method with that of \citet{Obreschkow09}}

\subsection{Method O09: model $\SHI$ analytically considering the HI-H$_2$ conversion} 
\subsubsection{Model 1: the original model of Obreschkow et al. 2009}
Motivated by the following results, Obreschkow et al. (2009) (O09 hereafter) proposed an analytical model for the radial distribution of $\SHI$.
\begin{enumerate}
\item Both the surface density of total neutral gas ($\hi+\htwo$) and the gas conversion ratio ($\htwo/\hi$) in galaxies are observed to follow radial profiles that are close to exponential functions \citep{Leroy08}. 
\item The localized $\htwo/\hi$ in galaxies are predicted to depend on the mid-plane pressure, $P$, contributed mostly by the stars and the neutral gas. The correlation between $\htwo/\hi$ and $P$ has been confirmed by observations \citep{Leroy08}. 
\end{enumerate}

The original O09 model has the function form
\begin{equation}
  \SHI(r)  =  \frac{\Sigmastd \exp(-r/\rd)}{1+\fc\exp(-1.6 r/\rd)},     
 \label{eq:SigHI_O09}
\end{equation}

\begin{equation}
  \fc  =   [K~\rd^{-4} M_{\rm gas} (M_{\rm gas}+ \langle f_{\sigma} \rangle M_*)]^{0.8},    
 \label{eq:Rmol_O09} 
\end{equation}

where $\Sigmastd$ is the central gas surface density, $\rd$ is the scale-length of the gas profile, $\fc$ represents the central $\htwo/\hi$ ratio, $M_{\rm gas}=1.36(\mHI+M_{\rm H_2}$), $K \equiv 11.3 {\rm m^4~ kg^{-2} }$, and  $\langle f_{\sigma} \rangle \sim0.4$.

\citet{Wang14} used a similar model as Eq.~\ref{eq:SigHI_O09}, but treated $\Sigmastd$ and $\fc$ as free parameters, to successfully fit the observed $\SHI$ profiles of 39 galaxies. $\Sigmastd$ is also adjustable in O09, but many assumptions have been made to produce Eq.~\ref{eq:Rmol_O09}, including that $\rd$ is twice the scale-length of the stellar disk ($r_{\rm s}$), and that the velocity dispersion is constant for the gas and exponentially rising as a function of radius for the stars.

Because $\hi$ is on average more abundant and more extended than the $\htwo$ gas, we further assume $\rd$ to be the scale-length of the $\hi$ profile, so $\rd \sim 0.2\rHI$ \citep[W16,][]{Wang14}. 
We hence can use the O09 model to guess the radial distribution of \hi, $\SHI$ for ES2, based on the integrated $\mHI$, $\mHtwo$ and $M_*$. The corresponding $\hi$ mass within $r_{90}$, $M_{HI,in,2}$, can also be calculated.

\subsubsection{Modified O09 models}
There are a few possible modifications to the original O09 model.

For the majority of galaxies from xGASS or ES2, which do not have $\htwo$ observations, we can approximate $M_{\rm H_2}=0.2 \mHI$, based on the average $\htwo/\hi$ of galaxies with $M_*>10^9 M_{\odot}$ \citep{Saintonge11}. We note that $\htwo/\hi$ increases as a function of $M_*$ \citep{Catinella18}, however as we will show later that using accurate $\htwo/\hi$ does not significantly improve the method.

$\rd$ need not necessarily be assumed to be $2r_{\rm s}$, as it can be approximated as 0.2$\rHI$ (W16). Then, using a similar deduction procedure as in O09, we obtain the following estimate of $\fc$ that can be used to replace Eq.~\ref{eq:Rmol_O09}:
\begin{equation}
\fc=  (\Sigma_{\rm gas} (\Sigma_{\rm gas}+\Sigma_{\rm *,eff}))^{0.8} K^{0.8}  ,
\label{eq:Rmol_here} 
\end{equation}

where 
\begin{eqnarray}
\Sigma_{\rm gas} &=&  M_{\rm gas}/\rd^2 \exp(-r/\rd), \\
\Sigma_{\rm *,eff} &=&  \langle f_{\sigma} \rangle/4 M_*/r_{\rm s}^2 \exp(-r/r_{\rm s}).
\end{eqnarray}

$\Sigma_{\rm *,eff}$ can be viewed as the surface density of stars which effectively contribute to the mid-plane pressure, and hence to $\fc$.

We predict $\SHI$ and $\mHIin$ with the following modified O09 models for the galaxies from ES2:
\begin{enumerate}
\item {\bf Model 2:} $M_{\rm H_2}$ is replaced by $0.2 \mHI$, and $\fc$ is estimated with Eq.~\ref{eq:Rmol_O09}. This model requires only input of $\mHI$ and $M_*$, hence the minimum number of input parameters among the different models. 

\item {\bf Model 3:} $\mHtwo$ unchanged, and $\fc$ is estimated with Eq.~\ref{eq:Rmol_here}. This model uses real measurements of $M_{\rm gas}$ and $r_{\rm s}$, hence relies on less assumptions than the other three models. 

\item {\bf Model 4:} $M_{\rm H_2}$ is replaced by $0.2 \mHI$, and $\fc$ is estimated with Eq.~\ref{eq:Rmol_here}. This model requires to input $\mHI$, $M_*$ and $r_{\rm s}$, which are in principle available for $\hi$ surveys like xGASS, which has optical images from SDSS. 
\end{enumerate}

\subsection{Selection among the methods}
Like in Sec.~\ref{sec:method1}, we compare the estimated $\mHIin$ with the real measurements $\mHIin0$ to assess the different methods. 
In addition to VS, we select the 10 THINGS galaxies from VS and call them VS2. These galaxies have CO (and hence the derived $\htwo$) images from BIMA SONG \citep{Helfer03} or HERACLES \citep{Leroy09}. Scale-length of the stellar disks ($r_{\rm s}$) have been measured in \citet{Leroy08}. The availability of $\htwo$ images and $r_{\rm s}$ makes it possible the applications of model 1, 3 and 4 of method O09.

We apply method W16 (which was presented in Sec.\label{sec:method1}) and the different models of method O09 to ES2. We quantify the difference between the real and predicted amount of $\hi$ within $r_{90}$, $\log~M_{\rm HI,in,pred}/M_{\rm HI,in}$ and $\Sigma_{\rm HI,in,pred}-\Sigma_{\rm HI,in}$. The median and scatter (standard deviation) of the differences in ES2 with different methods are listed in Tab.~\ref{tab:test_ES2}. We note that the median absolute differences are less important than the scatters when assessing the performance of methods, for they can be calibrated (with ES2) and systematically removed later.  

We firstly compare between the four models of method O09.
Model 1 and 2 have smaller scatter, and hence work better than model 3 and 4. Model 2 uses the minimum number of input parameters, only $\mHI$ and $M_*$, hence the inclusion of $\mHtwo$ or $r_{\rm s}$ in the other three models seems not significantly improving the predictions. It implies that due to the complexity of the  $\hi$-to-$\htwo$ process, the uncertainties of the models are large compared to the uncertainties of the input parameters. 
This is good news because $\mHtwo$ are only available for part of the xGASS sample and accurate measurements of $r_{\rm s}$ are tricky due to the contamination of disk breaks, bars and bulges \citep{Gao17}. We hence choose model 2 among the four models of method 2 for the remaining analysis of this section, for it performs better than model 3 and 4, and requires fewer input parameters than model 1. 

Within VS2, method W16 produces slightly larger scatter in $\log~M_{\rm HI,in,pred}/M_{\rm HI,in}$ and $\Sigma_{\rm HI,in,pred}-\Sigma_{\rm HI,in}$ than model 2 of method O09. Considering the relatively small sample size of VS2, we further use VS to compare between method W16 and model 2. We have added -0.04 and -0.08 dex to the direct estimates of method W16 and model 2 respectively, to minimize the scatter and median offset from real measurements in VS. 

The results of the comparison are displayed in Tab.~\ref{tab:test_ES}. Method W16 produces slightly larger scatter in $\log~M_{\rm HI,in,pred}/M_{\rm HI,in}$, but slightly smaller scatter in $\Sigma_{\rm HI,in,pred}-\Sigma_{\rm HI,in}$ than Method O09.  We further find with a figure close to Fig.~~\ref{fig:dmhi_correlate}, that both methods produce $\log~M_{\rm HI, in,pred}/M_{\rm HI, in}$ that do not significantly depend on $M_*$, $\mHI/M_*$ or $sSFR$. Putting these comparisons together, the two methods have similar performances in estimating $\mHIin$ and $\SHIin$. {\bf We choose method W16 for its simplicity (only requiring $\mHI$ as input, and model independent), for the analysis in this paper.}

\begin{figure*}
\centering
\includegraphics[width=5.5cm,angle=0]{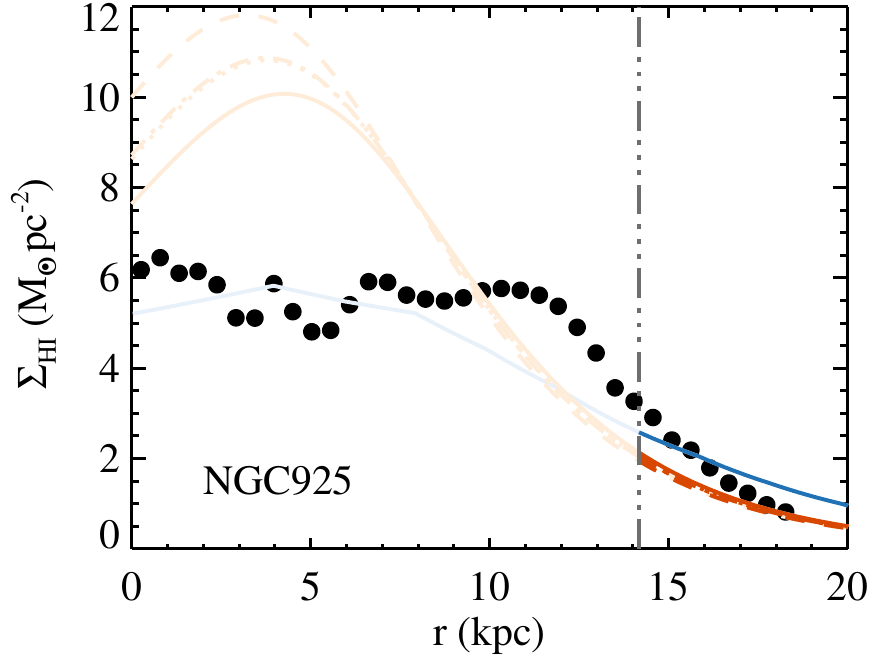}
\includegraphics[width=5.5cm,angle=0]{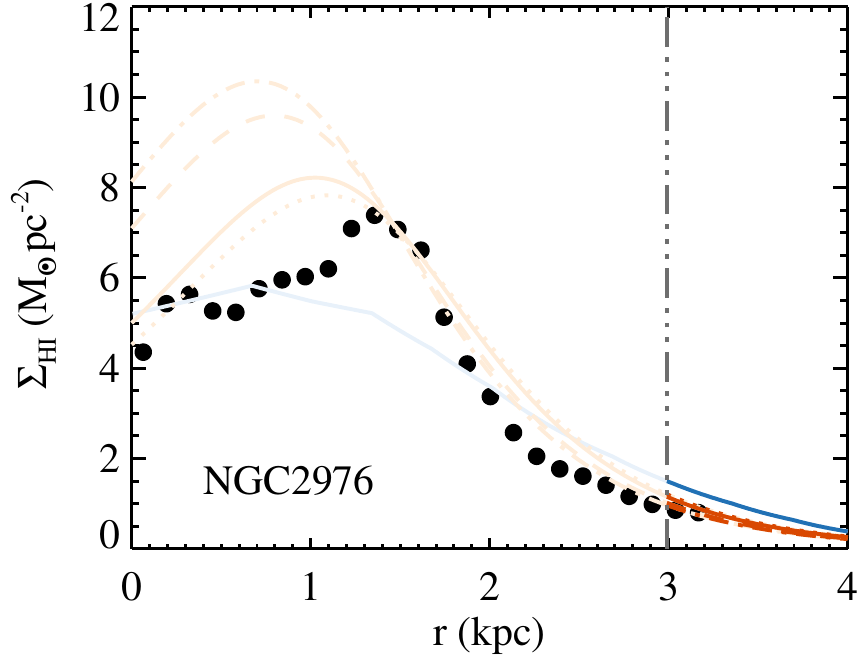}
\includegraphics[width=5.5cm,angle=0]{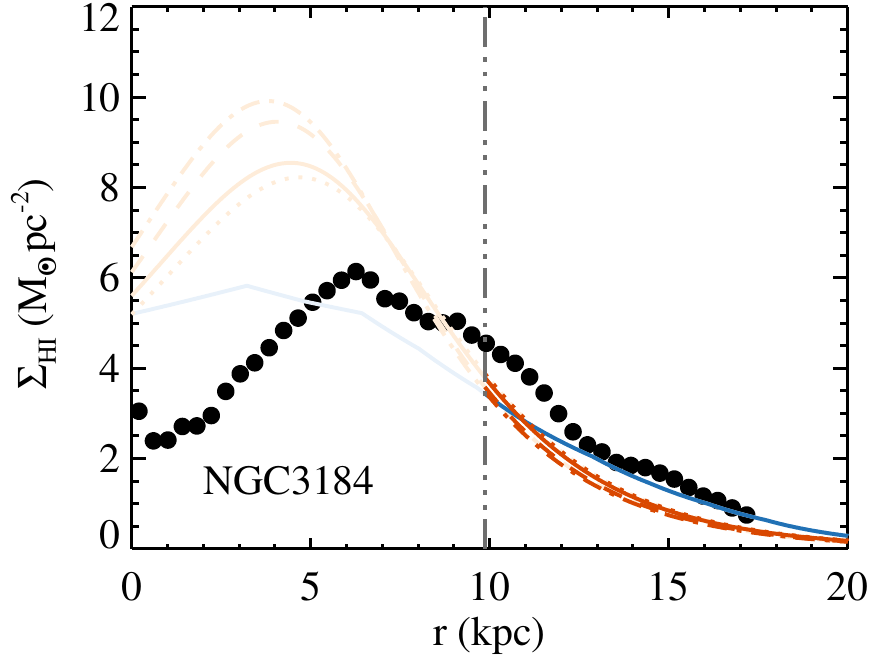}

\includegraphics[width=5.5cm,angle=0]{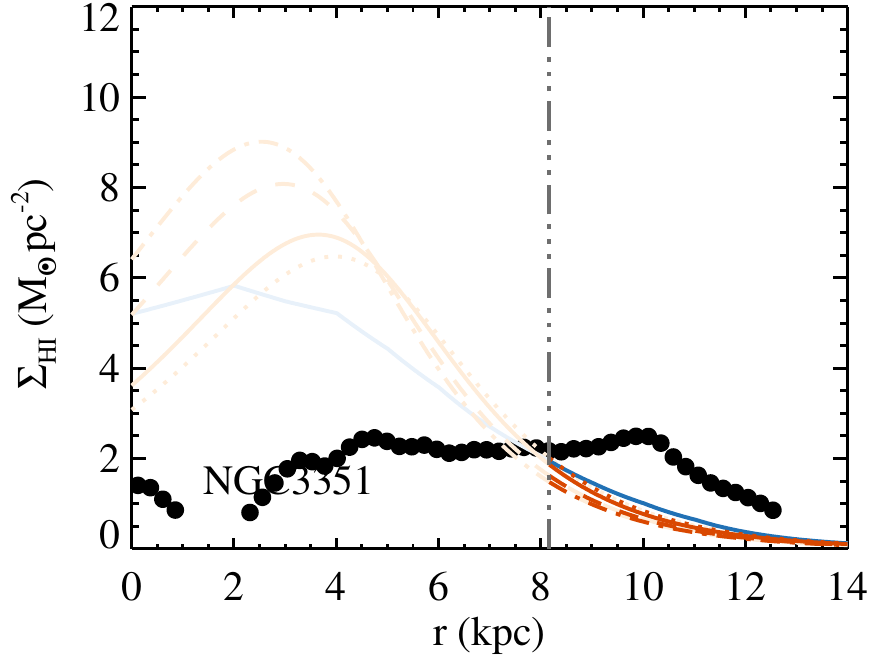}
\includegraphics[width=5.5cm,angle=0]{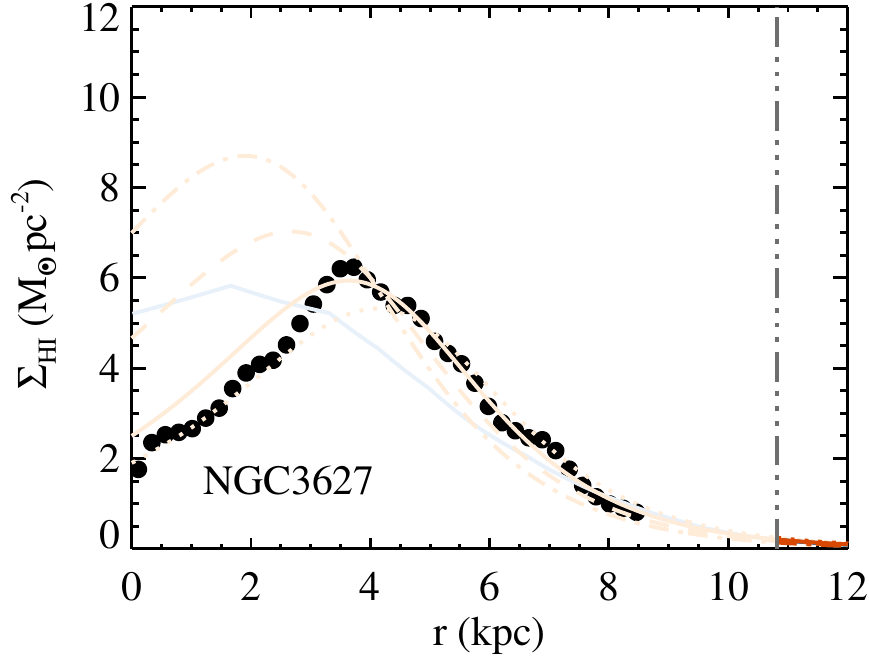}
\includegraphics[width=5.5cm,angle=0]{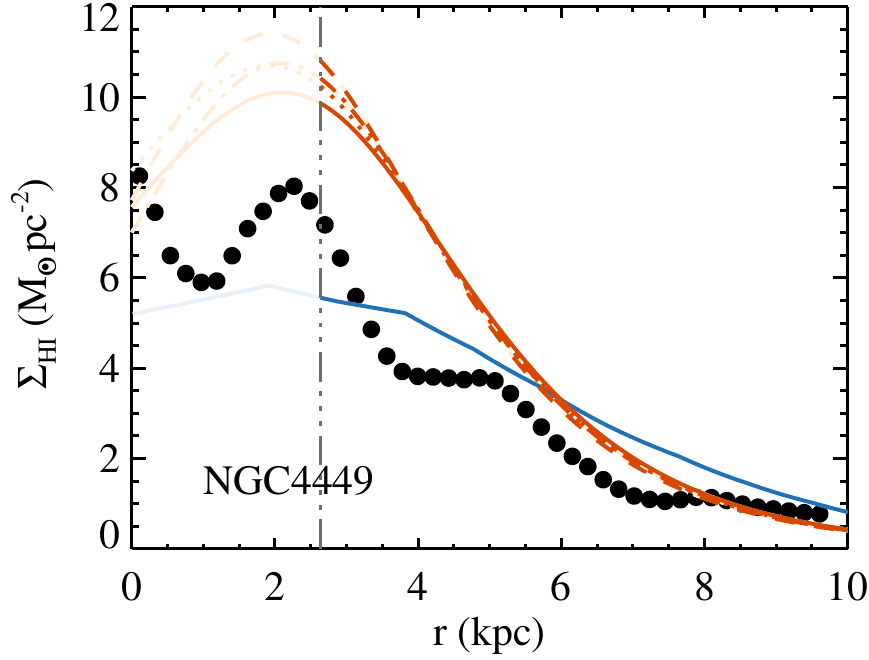}

\includegraphics[width=5.5cm,angle=0]{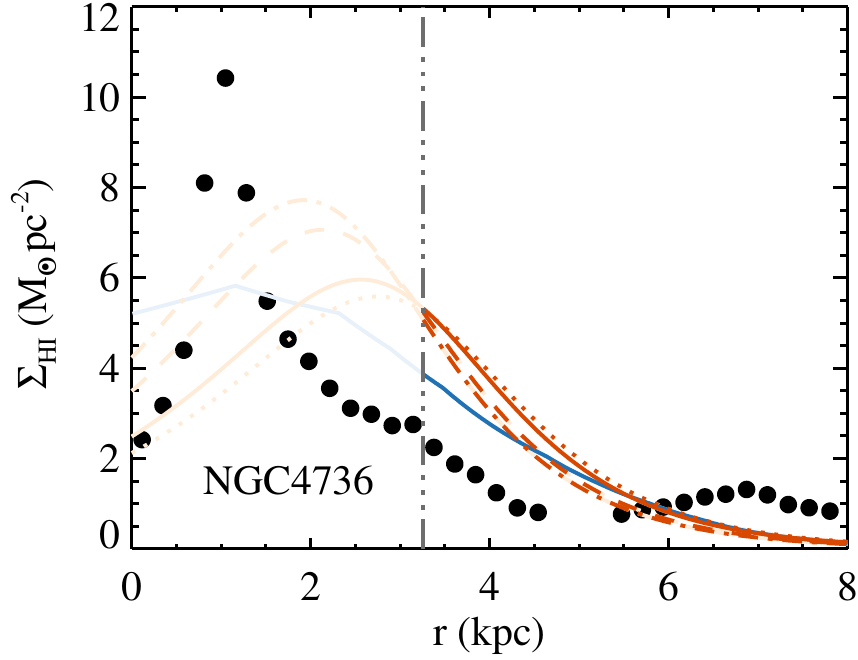}
\includegraphics[width=5.5cm,angle=0]{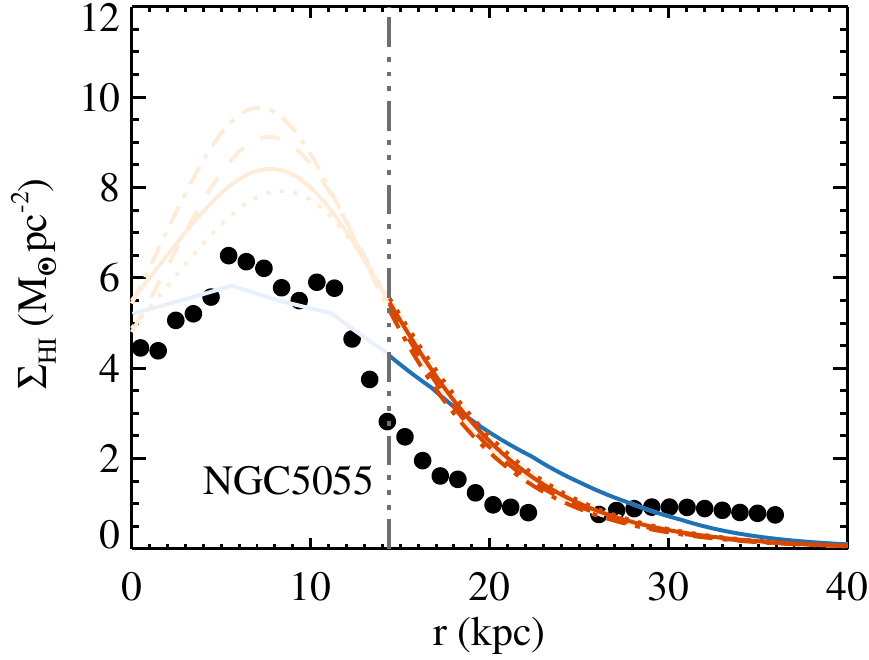}
\includegraphics[width=5.5cm,angle=0]{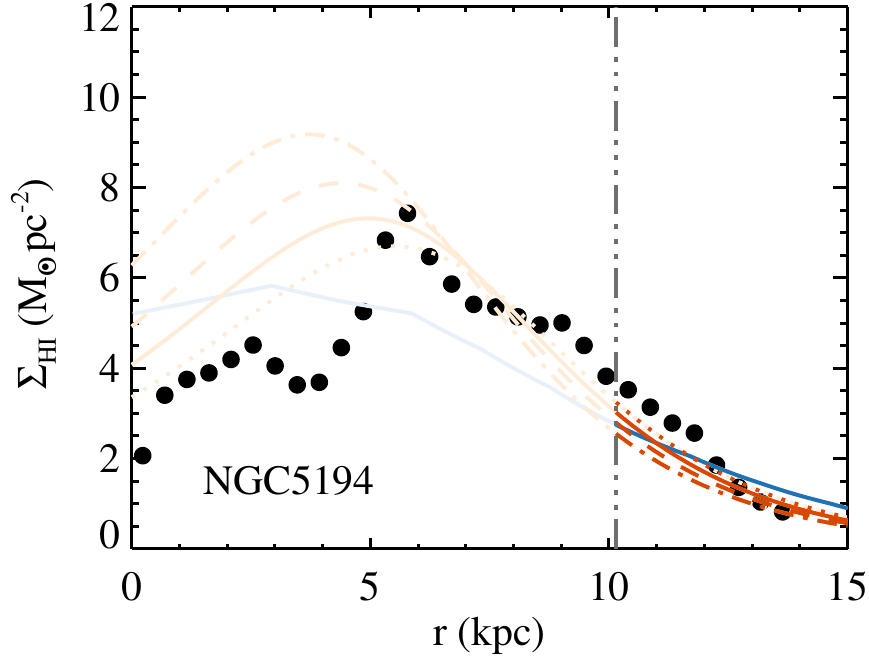}

\includegraphics[width=5.5cm,angle=0]{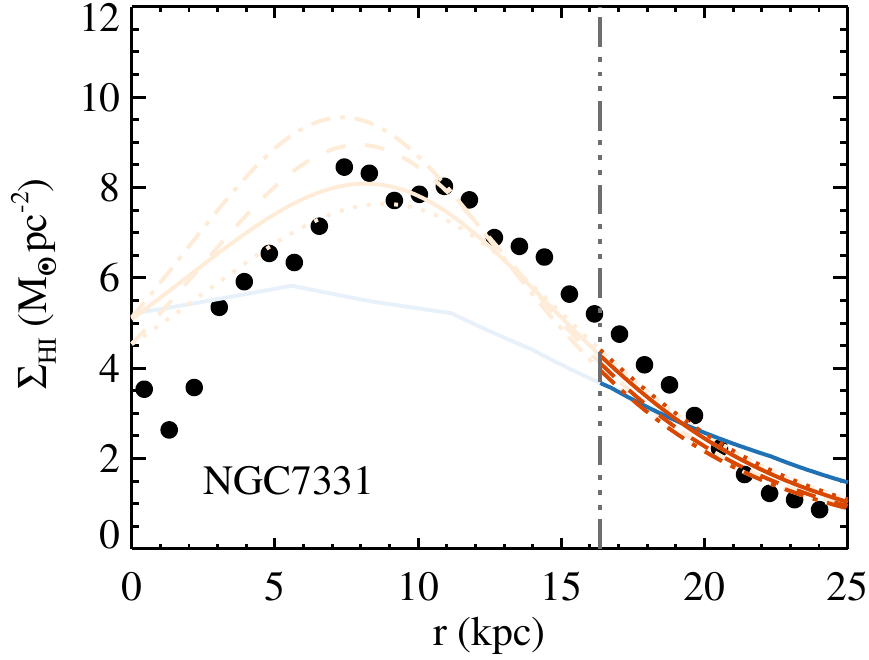}
\includegraphics[width=5.5cm,angle=0]{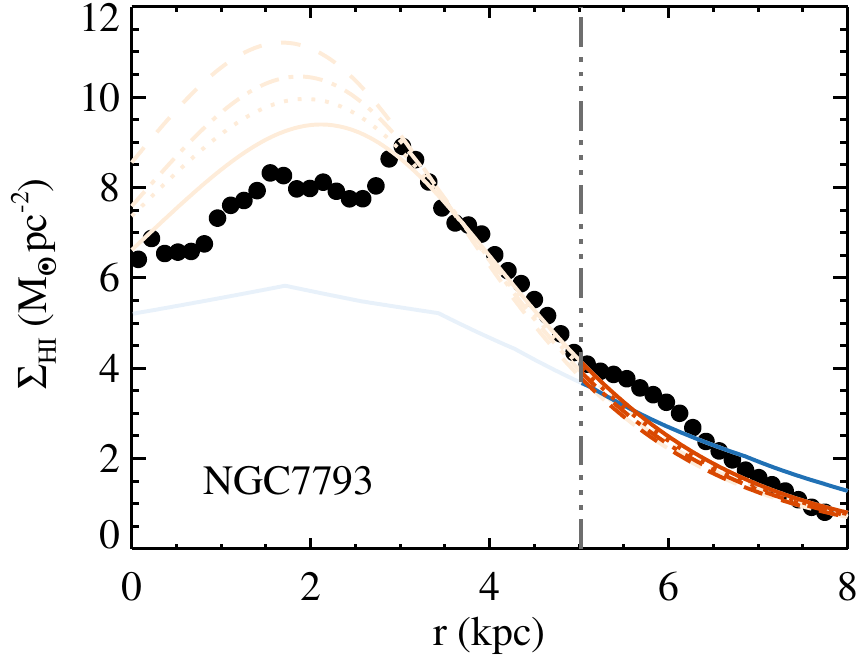}

\caption{The real and predicted $\hi$ radial distributions. The black dots show the real measurements. The blue curves are predictions from method 1. The orange dotted, solid, dashed, and dot-dashed curves are predictions from model 1-4 of method 2 respectively. The black dash-three-dotted lines mark the position of optical $r_{90}$. }
\label{fig:things_prof}
\end{figure*}

In case the readers might be interested, Fig.~\ref{fig:things_prof} compares the predicted $\hi$ radial distributions to the real measurements in ES2. The models of method O09 match the real profiles in a remarkably close way in the outer regions, but much less closely in the inner regions. If we compare among the models of method O09, model 2 (orange solid curve) seems to provide the closest match to the $\SHI$ profiles within $r_{90}$, though physically it appears to be the most simplified model. The predicted profiles of method W16  never match the inner or outer profiles perfectly, but has been able to provide a reasonably close estimates to the $\hi$ mass and average surface densities in the inner regions.

\begin{table}
\centering
\caption{Comparing predicted $\mHIin$ and $\SHIin$ with real measurements in VS2}
{
\begin{tabular}{c|cc|cc}
Method &  \multicolumn{2}{|c|}{$\log M_{\rm HI,in,pred}/\mHIin$} & \multicolumn{2}{|c}{$\Sigma_{\rm HI, in, pred}-\SHIin$} \\
  &  \multicolumn{2}{|c|}{} &  \multicolumn{2}{|c}{($M_{\odot} pc^{-2}$)} \\
  &   median & $\sigma$  & median & $\sigma$ \\
\hline
W16		 & 0.06   &     0.09    &    0.82    &    1.37 \\
\hline
O09-model 1 &0.09   &     0.05   &     1.11      &  1.15 \\
O09-model 2 &  0.10     &   0.05  &      1.14     &   1.10 \\
O09-model 3 &  0.14  &      0.05   &     1.59      &  1.21 \\
O09-model 4 & 0.14    &    0.06  &      1.70   &     1.20 \\
\end{tabular}
}
\label{tab:test_ES2}
\end{table}

\begin{table}
\centering
\caption{Comparing predicted $\mHIin$ and $\SHIin$ with real measurements in VS}
{
\begin{tabular}{c|cc|cc}
Method &  \multicolumn{2}{|c|}{$\log M_{\rm HI,in,pred}/\mHIin$} & \multicolumn{2}{|c}{$\Sigma_{\rm HI, in, pred}-\SHIin$} \\
  &  \multicolumn{2}{|c|}{} &  \multicolumn{2}{|c|}{ ($M_{\odot} pc^{-2}$) } \\
  &   median & $\sigma$  & median & $\sigma$ \\
\hline
W16		& -0.01  &  0.09 &  -0.04  &  0.60 \\
O09-model 2 & 0.00  &  0.07  &  -0.01 &   0.81 \\
\end{tabular}
}

\vspace{0.5cm}
\label{tab:test_ES}
\end{table}

\section{2. Trends of $\SHIin$ when the inner disks are defined by $R_{25}$ }
\begin{figure*}
\centering
\includegraphics[width=8.8cm,angle=0]{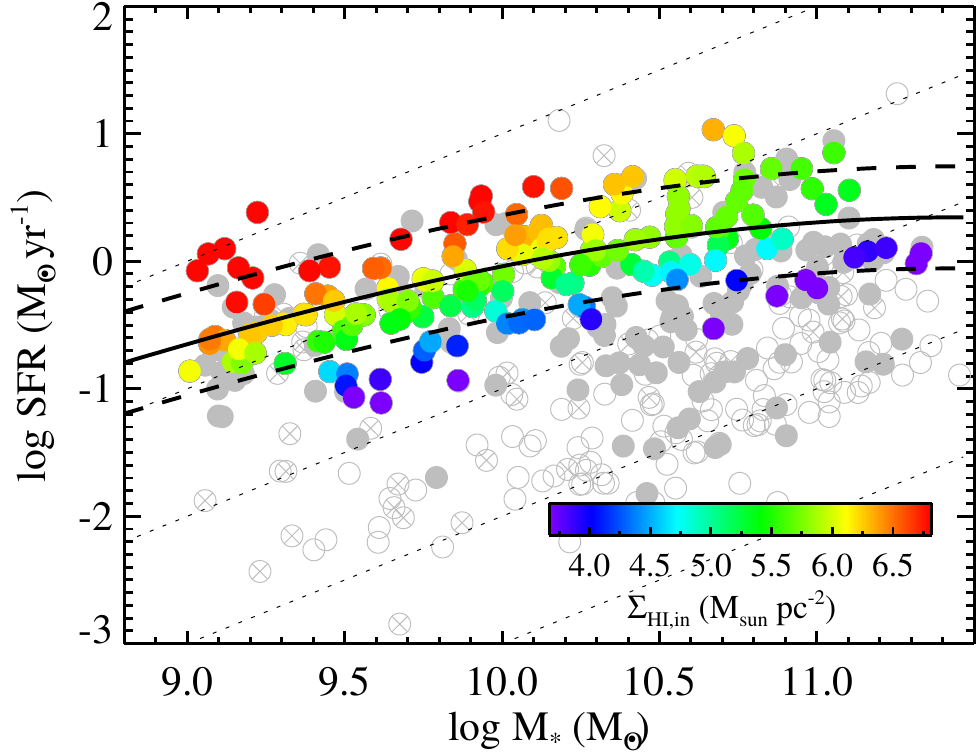}
\includegraphics[width=8.8cm,angle=0]{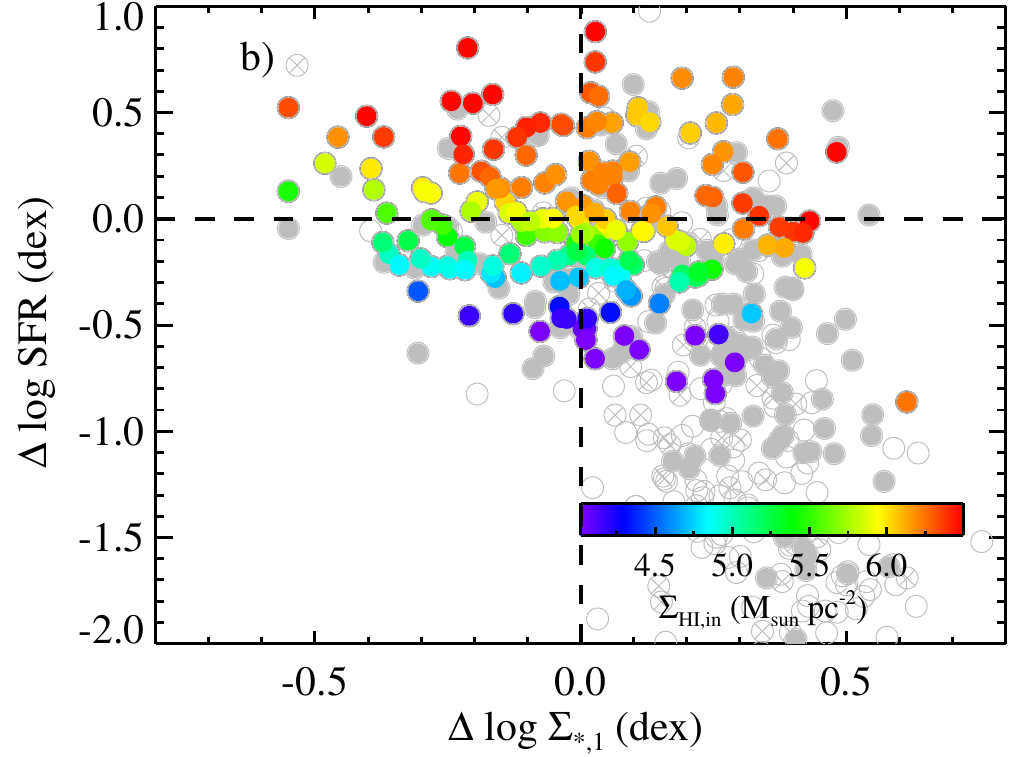}
\vspace{0.2cm}
\caption{The left and right panels are similar to panel e of Fig.~\ref{fig:SFMS} and panel d of Fig.~\ref{fig:dsfdsz} respectively. The only difference is that $R_{25}$ instead of $r_{90}$ are used to estimate $\mHIin$. }
\label{fig:dsfdsz_R25}
\end{figure*}

 There might be worry that using $r_{90}$ in the estimates of $\mHIin$ may cause systematic uncertainties due to the dependence of $r_{90}$ on the significance of bulges. We hence test by replacing $r_{90}$ with $R_{25}$ in the estimate of $\mHIin$, but find all trends presented in the main part of this paper remain. We show two example plots in Fig.~\ref{fig:dsfdsz_R25}, which are analogs of panel e of Fig.~\ref{fig:SFMS} and panel d of Fig.~\ref{fig:dsfdsz}, but with the new $\mHIin$ estimates. We can see that the trends are similar with the two types of $\mHIin$ estimates.

\end{document}